\documentclass[twocolumn]{svjour3}          

\smartqed  

\usepackage{graphicx}

\usepackage{amssymb,amsmath} 
\usepackage{graphics}
\usepackage{epsfig}
\usepackage{algorithm}
\usepackage{algpseudocode}
\usepackage{color}
\usepackage{natbib}
\usepackage{multirow}

\usepackage[english]{babel}

\newcommand{\dfn}{\triangleq}

\newcommand{\rw}{\rightarrow}

\newcommand{\sS}{{\sf S}}

\newcommand{\mB}{{\mathcal B}}

\newcommand{\mX}{{\mathcal X}}
\newcommand{\mY}{{\mathcal Y}}
\newcommand{\Real}{\mathbb{R}}
\newcommand{\mbN}{\mathbb{N}}
\newcommand{\mbE}{\mathbb{E}}
\newcommand{\mbP}{\mathbb{P}}
\newcommand{\bftheta}{{\boldsymbol{\theta}}}

\begin{document}

\title{A comparison of nonlinear population Monte Carlo and particle Markov chain Monte
Carlo algorithms for Bayesian inference in stochastic kinetic models
\thanks{
E. K.  acknowledges the support of \textit{Ministerio de
Educaci\'on} of Spain (\textit{Programa de Formaci\'on de
Profesorado Universitario}, ref. AP2008-00469). This work has been
partially supported by {\em Ministerio de Econom\'{\i}a y
Competitividad} of Spain (program Consolider-Ingenio 2010
CSD2008-00010 COMONSENS and project COMPREHENSION
TEC2012-38883-C02-01). }}

\titlerunning{Population Monte Carlo vs Markov chain Monte Carlo}        

\author{Eugenia Koblents         \and
        Joaqu\'{\i}n M\'{\i}guez }


\institute{E. Koblents and J. M\'{\i}guez \at
              Department of Signal Theory and Communications, Universidad Carlos III de Madrid, Madrid, Spain \\
              \email{ekoblents,jmiguez@tsc.uc3m.es}           
}

\date{Received: date / Accepted: date}

\maketitle

\begin{abstract}

In this paper we address the problem of Monte Carlo approximation of
posterior probability distributions in stochastic kinetic models
(SKMs). SKMs are multivariate Markov jump processes that model the
interactions among species in biochemical systems according to a set
of uncertain parameters. Markov chain Monte Carlo (MCMC) methods
have been typically preferred for this Bayesian inference problem.
Specifically, the particle MCMC (pMCMC) method has been recently
shown to be an effective, while computationally demanding, method
applicable to this problem. Within the pMCMC framework, importance
sampling (IS) has been used only as the basis of the sequential
Monte Carlo (SMC) approximation of the acceptance ratio in the
Metropolis-Hastings kernel. However, the recently proposed nonlinear
population Monte Carlo (NPMC) algorithm, based on an iterative IS
scheme, has also been shown to be effective as a Bayesian inference
tool for low dimensional (predator-prey) SKMs. In this paper, we
provide an extensive performance comparison of pMCMC versus NPMC,
when applied to the challenging prokaryotic autoregulatory network.
We show how the NPMC method can greatly outperform the pMCMC
algorithm in this scenario, with an overall moderate computational
effort. We complement the numerical comparison of the two techniques
with an asymptotic convergence analysis of the nonlinear IS scheme
at the core of the proposed method when the importance weights can
only be computed approximately.


\end{abstract}

\section{Introduction}
\label{Introduction}


Stochastic kinetic models (SKMs) are multivariate systems that model
molecular interactions among species in biological and chemical
problems, according to a set of unknown rate parameters
\citep{Wilkinson2011}. The aim of this paper is the approximation of
the posterior distribution of the rate parameters and the
populations of all species, provided a set of discrete, noisy
observations is available.
This inference problem has been traditionally addressed using Markov
chain Monte Carlo (MCMC) schemes \citep{Boys08,
Milner2013,Wilkinson2011paper,Wilkinson2011}. In
\citep{Golightly2011} a particle MCMC (pMCMC) method
\citep{Andrieu2010} has been successfully applied to this problem.
The pMCMC technique relies on a sequential Monte Carlo (SMC)
approximation of the posterior distribution of the populations to
compute the Metropolis-Hastings (MH) acceptance ratio.


However, MCMC methods in general, and pMCMC in particular, suffer
from a number of problems. The convergence of the Markov chain is
hard to assess and the final set of samples presents correlations
which can greatly reduce its efficiency. Besides, MCMC methods do
not (easily) allow for parallel implementations and turn out to be
computationally intensive. To reduce the complexity of the existing
MCMC methods when applied to SKMs, a diffusion approximation of the
underlying stochastic process is usually applied
\citep{Golightly2005}. The parameters of the MCMC proposal are also
hard to choose and determine the performance of the algorithm.


An appealing alternative to the widely established MCMC methods is
the population Monte Carlo (PMC) algorithm \citep{Cappe04}. PMC is
an iterative importance sampling (IS) scheme that yields a discrete
approximation of a target probability distribution. The PMC
algorithm has important advantages with respect to MCMC techniques.
It provides independent samples and asymptotically unbiased
estimates at all iterations, which avoids the need of a convergence
period. Additionally, PMC may be easily parallelized.

On the other hand, the main weakness of IS and PMC is their low
efficiency in high dimensional problems, due to the well known
degeneracy problem \citep{Bengtsson08}. The recently proposed
nonlinear PMC (NPMC) scheme \citep{Koblents2013a} mitigates this
difficulty by computing nonlinear transformations of the importance
weights (IWs), in order to smooth their variations and avoid
degeneracy. In \citep{Koblents2013a} a simple convergence analysis
of nonlinear IS (NIS) is provided, for two types of nonlinear
transformations, \emph{tempering} and \emph{clipping}. Similarly to
the pMCMC method in \citep{Golightly2011}, the NPMC method resorts
to an SMC approximation of the posterior distribution of the
populations to compute, in our case, the IWs.

In \citep{Koblents2013b,Koblents2013c} the nonlinear version of IS
and PMC is combined with the popular mixture-PMC (MPMC) method of
\citep{Cappe08}, which allows to approximate arbitrary
high-dimensional target distributions by means of mixtures of
Gaussian or t-Student distributions. The original MPMC algorithm of
\citep{Cappe08} has been applied to cosmological inference problems
and compared to an MCMC method in \citep{Wraith2009} (and
\citep{Kilbinger2010}), and has been shown to provide similar
precision results with a lower computation load than its MCMC
counterpart. The MPMC scheme is the basis of the tool CosmoPMC
\citep{Kilbinger2012} for the estimation of cosmological parameters,
as an alternative to the MCMC package, CosmoMC, \citep{Lewis2002}
http://cosmologist.info/cosmomc.

In this paper we apply the NPMC method to the estimation of both the
parameters and the unobserved populations in SKMs. We present
numerical results to compare the performance of the state-of-art
pMCMC and the proposed NPMC, when applied to the challenging
prokaryotic model in two scenarios of different dimension and with
two different observation models. We show that the NPMC method
outperforms the pMCMC method for the same computational cost.

As a complement to the numerical comparison, we introduce new
asymptotic convergence results for the NIS scheme that accounts for
the use of SMC to approximate the IWs. The analysis in this paper
considerably extends the preliminary results in
\citep{Koblents2013a}. In particular, we prove that approximate
integrals computed via NIS converge almost surely (as the number of
samples increases) and explicit convergence rates are given.


The rest of the paper is organized as follows. In Section \ref{SKMs}
we present an introduction to the basics of SKMs and the usual
solutions to this Bayesian inference problem. In Sections
\ref{pMCMC_section} and \ref{PMC_section} we describe the pMCMC and
NPMC methods, respectively, when applied to the approximation of
posterior distributions in SKMs. In Section \ref{SKM_example} we
numerically compare the performance of pMCMC and NPMC schemes when
applied to a prokaryotic autoregulatory model, with different
simulation settings. Section \ref{Analysis} is devoted to the
convergence analysis of the NIS method. Finally, Section
\ref{Conclusion} is devoted to the conclusions.

\section{Bayesian inference for stochastic kinetic models}
\label{SKMs}

\subsection{Stochastic kinetic models}

A SKM is a multivariate continuous-time jump process modeling the
interactions among molecules, or species, that take place in
chemical reaction networks of biochemical and cellular systems
\citep{Wilkinson2011}.

Consider a biochemical reaction network that describes the time
evolution of the population of $V$ species $x_1, \ldots, x_V$
related by means of $K$ reactions $r_1, \ldots, r_K$ 
\begin{eqnarray*}
\begin{array}{cc}
  r_1 : & p_{11} x_1 + p_{12} x_2 + \ldots + p_{1V} x_V \stackrel{c_1} {\longrightarrow} \\
   & q_{11} x_1 + q_{12} x_2 + \ldots + q_{1V} x_V, \\
  r_2 : & p_{21} x_1 + p_{22} x_2 + \ldots + p_{2V} x_V \stackrel{c_2} {\longrightarrow} \\
  & q_{21} x_1 + q_{22} x_2 + \ldots + q_{2V} x_V, \\
  \vdots & \vdots \\
  r_K : & p_{K1} x_1 + p_{K2} x_2 + \ldots + p_{KV} x_V \stackrel{c_K} {\longrightarrow} \\
  & q_{K1} x_1 + q_{K2} x_2 + \ldots + q_{KV} x_V, \\
\end{array}
\end{eqnarray*}
where $p_{kv}$ and $q_{kv}$, $k=1,\ldots,K$, $v=1,\ldots,V$, denote
the reactant and the product coefficients, respectively; and $c_k >
0$, $k=1,\ldots,K$, are the random constant rate parameters. A
matrix $\textbf{P}$ of size $K \times V$ contains the reactant
coefficients $p_{kv}$ and, similarly, $\textbf{Q}$ contains the
product coefficients $q_{kv}$. The stoichiometry matrix of size $V
\times K$ is defined as $\textbf{S} = (\textbf{Q} -
\textbf{P})^\top$. The vector $\textbf{c} = [c_1, \ldots, c_K]^\top$
contains the rate parameters.

Let $x_v(t)$, $v=1,\ldots,V$, denote the nonnegative, integer
population of species $x_v$ at time $t$, and let $\textbf{x}(t) =
[x_1(t), \ldots, x_V(t)]^\top$ denote the state of the system at
this time instant. Let $\textbf{x}_n = [x_{1,n}, \ldots,
x_{V,n}]^\top$ denote the state of the system at discrete time
instants $t = n\Delta$, $n=1,\ldots,N$, i.e., $x_{v,n} =
x_v(n\Delta)$ where $\Delta$ denotes a time-discretization period.
We denote by $\textbf{x}$ the $VN \times 1$ vector containing the
population of each species at $N$ consecutive discrete time
instants, i.e., $\textbf{x} = [ \textbf{x}_1^\top, \ldots,
\textbf{x}_N^\top ]^\top$.

The $k$-th reaction takes place stochastically according to its
instantaneous rate or hazard function
\begin{equation*}
h_k(t) = c_k \prod_{v=1}^V { x_v(t) \choose p_{kv} }, \quad
k=1,\ldots,K,
\end{equation*}
where the product of binomial coefficients represents the number of
combinations in which the $k$-th reaction can occur, as a function
of the population of each reactant species $x_v$. We additionally
define the vector $\textbf{h}(t) = [h_1(t), \ldots,h_K(t)]^\top$.
The waiting time to the next reaction is exponentially distributed
with parameter $h_0(t) = \sum_{k=1}^K h_k(t)$, and the probability
of each reaction type is given by $h_k(t)/h_0(t)$.

\subsection{Bayesian inference for SKMs}

We consider the log-transformed rate parameters $\bftheta =
[\theta_1,\ldots,\theta_K]^\top$, where $\theta_k = \log(c_k)$,
$k=1,\ldots,K$, with prior pdf $p (\bftheta)$. The prior pdf of the
initial population vector $\textbf{x}_0$ is denoted by
$p(\textbf{x}_0)$. We assume that a linear combination of the
populations of a subset of species is observed at discrete time
instants corrupted by Gaussian noise, i.e.,
\begin{equation}
\label{eq_LH} \textbf{y}_n = \textbf{M} \textbf{x}_n + \textbf{w}_n,
\quad n=1,\ldots,N,
\end{equation}
where $\textbf{M}$ is the observation matrix with dimensions $D
\times V$ and $\textbf{w}_n \sim \mathcal{N}_D(\textbf{w}_n;
\textbf{0}, \sigma^2 \textbf{I})$  is a multivariate Gaussian noise
component. We denote the complete observation vector with dimension
$DN \times 1$ as $\textbf{y} = [\textbf{y}_1^\top, \ldots,
\textbf{y}_N^\top]^\top$.

The dynamical behavior of an arbitrary SKM may be described in terms
of the following set of equations\footnote{For simplicity of
notation, in this section we use $p$ to denote the pdfs in the
model. We write conditional pdfs as $p(\textbf{y}|\textbf{x})$, and
joint densities as $p(\bftheta) = p(\theta_1, \ldots, \theta_K)$.
This is an argument-wise notation, hence $p(\theta_1)$
denotes the distribution of $\theta_1$, possibly different from $p(\theta_2)$.} 
\begin{equation*}
\left\{ \begin{array}{ll}
          \bftheta \sim p(\bftheta) & \quad \textrm{(parameters prior)}, \\
          \textbf{x}_0 \sim p(\textbf{x}_0) & \quad \textrm{(populations prior)}, \\
          \textbf{x}_n \sim p (\textbf{x}_n | \textbf{x}_{n-1}, \bftheta) & \quad \textrm{(transition equation)}, \\
          \textbf{y}_n \sim p(\textbf{y}_n | \textbf{x}_n) & \quad \textrm{(observation equation),}
        \end{array}
 \right.
\end{equation*}
where $p (\textbf{x}_n | \textbf{x}_{n-1},\bftheta)$ and
$p(\textbf{y}_n | \textbf{x}_n)$ denote the transition pdf and the
likelihood function, respectively. The Gillespie algorithm
\citep{Gillespie77} allows to perform exact forward simulations of
arbitrary SKMs, drawing samples from the transition densities
$p(\textbf{x}_n | \textbf{x}_{n-1}, \bftheta)$, $n=1,\ldots,N$,
given a set of log-rate parameters $\bftheta$ and an initial
population $\textbf{x}_0$.

In this paper, we aim to obtain a Monte Carlo approximation of the
full joint posterior distribution of the log-rate parameters
$\bftheta$ and the populations $\textbf{x}$, with density
\begin{equation}
\label{eq_post} p (\bftheta, \textbf{x} | \textbf{y}) \propto p
(\textbf{y} | \textbf{x} ) p (\textbf{x} | \textbf{x}_0, \bftheta) p
(\textbf{x}_0) p (\bftheta),
\end{equation}
given the prior distributions $p(\bftheta)$ and $p(\textbf{x}_0)$,
the transition pdf $p (\textbf{x} | \textbf{x}_0, \bftheta) =
\prod_{n=1}^N p(\textbf{x}_n | \textbf{x}_{n-1}, \bftheta)$ and the
likelihood function $p(\textbf{y} | \textbf{x}) = \prod_{n=1}^N
p(\textbf{y}_n | \textbf{x}_n)$ constructed from equation
(\ref{eq_LH}).

We are also interested in computing approximations of the posterior
marginals of the rate parameters $p(\bftheta | \textbf{y}) = \int p
(\bftheta, \textbf{x} | \textbf{y}) d\textbf{x}$ and the species
populations $p(\textbf{x} | \textbf{y}) = \int p (\bftheta,
\textbf{x} | \textbf{y}) d \bftheta$ as well as their moments (e.g.,
the posterior mean), which are of the form
\begin{equation*}
E_{p (\bftheta | \textbf{y})} [f(\bftheta)] = \int f(\bftheta) p
(\bftheta | \textbf{y}) d \bftheta, \; \textrm{and}
\end{equation*}
\begin{equation*}
E_{p (\textbf{x} | \textbf{y})} [f(\textbf{x})] = \int f(\textbf{x})
p (\textbf{x} | \textbf{y}) d\textbf{x}, \; \textrm{respectively},
\end{equation*}
where $f$ is a real, integrable function.

Bayesian inference based on exact stochastic simulations from
$p(\textbf{x}_n |\textbf{x}_{n-1}, \bftheta)$ generated via the
Gillespie algorithm often becomes practically intractable even for
models of modest complexity \citep{Golightly2005}. Thus, it is very
common to resort to a continuous approximation of the underlying
stochastic process, which is known as the diffusion approximation.
The diffusion process that most closely matches the dynamics of the
associated Markov jump process, over an infinitesimal time interval
$(t, t + dt]$, is given by a stochastic differential equation known
as the chemical Langevin equation (CLE) \citep{Wilkinson2011} (pag
230)
\begin{equation*}
d\textbf{x}(t) = \textbf{S} \, \textbf{h}(t) dt + \sqrt{\textbf{S}
\, \textrm{diag} \{ \textbf{h}(t) \} \textbf{S}^\top}
d\textbf{w}(t),
\end{equation*}
driven by the $V \times 1$ dimensional Wiener process
$\textbf{w}(t)$. However, this approximation is known to be poor in
low concentration scenarios, and thus should be avoided for models
involving species with a very low population. Alternatively, in
\citep{Milner2013} the authors propose a solution based on a moment
closure approximation of the stochastic process.

This inference problem has been traditionally addressed using MCMC
methods, and IS based schemes have been avoided due to their
inefficiency in high dimensional spaces \citep{Wilkinson2011}. In
\citep{Boys08} various MCMC algorithms are evaluated in data-poor
scenarios. In \citep{Golightly2011} a likelihood-free pMCMC scheme
\citep{Andrieu2010} is applied to this problem. This method is, to
the best of our knowledge, the most powerful, yet computationally
expensive, method provided so far for this kind of applications.

In \citep{Koblents2013a} a NPMC scheme is proposed for the
approximation of the marginal posterior pdf $p(\bftheta |
\textbf{y})$, which is computationally competitive, since it
requires the processing of a low number of samples of $\bftheta$ to
obtain the approximation of the posterior. The performance of the
NPMC method is tested in a simple SKM known as predator-prey model
\citep{Volterra26}, providing excellent results with a low
computational cost.

In this paper we compare the performances of the pMCMC and the NPMC
methods in the approximation of the full joint posterior
$p(\bftheta, \textbf{x} | \textbf{y})$ in equation (\ref{eq_post}),
which allows to perform Bayesian inference for the rate parameters
$\bftheta$ and the full sample path $\textbf{x}$, including
unobserved components.

\section{Particle MCMC for SKMs}
\label{pMCMC_section}

The particle marginal Metropolis-Hastings (PMMH) algorithm is a
pMCMC method originally proposed in \citep{Andrieu2010} for Monte
Carlo sampling from the full posterior distribution $p(\bftheta,
\textbf{x} | \textbf{y})$. The PMMH scheme suggests a proposal
mechanism of the form $q(\bftheta^\star | \bftheta)
\hat{p}^J(\textbf{x}^\star | \textbf{y}, \bftheta^\star)$. A new
candidate in the parameter space, $\bftheta^\star$, is drawn from an
arbitrary proposal distribution $q(\bftheta^\star | \bftheta)$,
while the new candidate in the variable space, $\textbf{x}^\star$,
is generated using an approximation of the posterior marginal
$p(\textbf{x}^\star | \textbf{y}, \bftheta^\star)$ constructed by
means of an SMC algorithm (i.e., a particle filter) with $J$
particles and denoted $\hat{p}^J(\textbf{x}^\star | \textbf{y},
\bftheta^\star)$. The probability of accepting the proposed pair
$(\bftheta^\star, \textbf{x}^\star)$ is
\begin{equation*}
\min \left\{1, \frac{\hat{p}^J(\textbf{y} |
\bftheta^\star)p(\bftheta^\star)}{\hat{p}^J(\textbf{y} |
\bftheta)p(\bftheta)} \times \frac{q(\bftheta |
\bftheta^\star)}{q(\bftheta^\star | \bftheta)} \right\},
\end{equation*}
where $\hat{p}^J(\textbf{y} | \bftheta^\star)$ is an unbiased
approximation of the marginal likelihood of $\bftheta^\star$ (i.e.,
$p(\textbf{y} | \bftheta^\star)$), computed, again, by way of a
particle filter with $J$ particles. The PMMH algorithm is reproduced
in Table \ref{pMCMCalgorithm}, and the SMC approximations of
$p(\textbf{y} | \bftheta^*)$ and $p(\textbf{x}^* | \textbf{y},
\bftheta^*)$ are described in Appendix A. Full details can be found
in \citep{Andrieu2010}. Note that the forward simulation of the
stochastic process in the particle filter may be performed exactly
with the Gillespie algorithm, or using a diffusion approximation.

\begin{table}[h]
\caption{Particle MCMC algorithm targeting $p(\bftheta, \textbf{x} |
\textbf{y})$ \citep{Andrieu2010}.} \vspace{-0.3cm}
\linethickness{0.3mm} \line(1,0){240} \vspace{0.1cm}
\underline{\textbf{Initialization ($i = 0$):}}
\begin{enumerate}
\item Sample $\bftheta^{(0)} \sim p(\bftheta)$ and
\item run a SMC scheme targeting $p(\textbf{x} | \textbf{y},
\bftheta^{(0)})$. Draw $\textbf{x}^{(0)} \sim \hat{p}^J(\textbf{x} |
\textbf{y}, \bftheta^{(0)})$ from the SMC approximation and let
$\hat{p}^J(\textbf{y} | \bftheta^{(0)})$ denote the marginal
likelihood estimate.
\end{enumerate}

\underline{\textbf{Iteration ($i = 1, \ldots, I$):}}
\begin{enumerate}
\item Sample $\bftheta^\star \sim q(\cdot |
\bftheta^{(i-1)})$ and
\item run a SMC scheme targeting $p(\textbf{x} | \textbf{y},
\bftheta^\star)$. Draw $\textbf{x}^\star \sim \hat{p}^J(\textbf{x} |
\textbf{y}, \bftheta^\star)$, let $\hat{p}^J(\textbf{y} |
\bftheta^\star)$ denote the marginal likelihood estimate, and
\item with probability
\begin{equation*}
\min \left\{ 1, \frac{\hat{p}^J(\textbf{y} |
\bftheta^\star)p(\bftheta^\star)}{\hat{p}^J(\textbf{y} |
\bftheta^{(i-1)})p(\bftheta^{(i-1)})} \times
\frac{q(\bftheta^{(i-1)} | \bftheta^\star)}{q(\bftheta^\star |
\bftheta^{(i-1)})} \right\}
\end{equation*}
accept the move setting $\bftheta^{(i)} = \bftheta^\star$,
$\textbf{x}^{(i)} = \textbf{x}^\star$ and $\hat{p}^J(\textbf{y} |
\bftheta^{(i)}) = \hat{p}^J(\textbf{y} | \bftheta^\star)$. Otherwise
store the current values $\bftheta^{(i)} = \bftheta^{(i-1)}$,
$\textbf{x}^{(i)} = \textbf{x}^{(i-1)}$ and $\hat{p}^J(\textbf{y} |
\bftheta^{(i)}) = \hat{p}^J(\textbf{y} | \bftheta^{(i-1)})$.
\end{enumerate}
\label{pMCMCalgorithm} \vspace{-0.2cm} \linethickness{0.3mm}
\line(1,0){240}
\end{table}

In \citep{Golightly2011} the proposal is selected as a Gaussian
random walk $q(\bftheta^\star | \bftheta) = \mathcal{N}_K
(\bftheta^\star; \bftheta, \gamma^2)$, whose variance $\gamma^2$ has
to be tuned and partly determines the performance of the algorithm.

After removing the initial burn-in samples and thinning the output,
we obtain a Markov chain $\{ \bftheta^{(i)},
\textbf{x}^{(i)}\}_{i=1}^M$ with $M$ correlated samples. Then, we
may construct a sample approximation of the marginal posterior
distributions of the parameters $\bftheta$ and the populations
$\textbf{x}$, as
\begin{equation*}
\hat{p}^M(d\bftheta | \textbf{y}) = \frac{1}{M} \sum_{i=1}^M
\delta_{\bftheta^{(i)}}(d\bftheta) \;\; \textrm{and}
\end{equation*}
\begin{equation*}
\hat{p}^M(d\textbf{x} | \textbf{y}) = \frac{1}{M}\sum_{i=1}^M
\delta_{\textbf{x}^{(i)}} (d\textbf{x}),
\end{equation*}
respectively, where $\delta_{\bftheta^{(i)}}$ and
$\delta_{\textbf{x}^{(i)}}$ denote the unit delta measure centered
at $\bftheta^{(i)}$ and $\textbf{x}^{(i)}$, respectively. The
approximation of the full joint posterior is of the form
\begin{equation*}
\hat{p}^M(d\bftheta, d\textbf{x} | \textbf{y}) = \frac{1}{M}\sum_{i=1}^M
\delta_{(\bftheta^{(i)}, \textbf{x}^{(i)})} (d\bftheta,
d\textbf{x}).
\end{equation*}

\section{Nonlinear PMC for SKMs}
\label{PMC_section}

The PMC method \citep{Cappe04} is an iterative IS scheme that
generates a sequence of proposal pdf's $q_\ell(\cdot)$, $\ell = 1,
\ldots, L$, that approximate a target pdf $\pi$ along the
iterations. In \citep{Koblents2013a} the NPMC scheme is proposed,
which introduces nonlinearly transformed IWs (TIWs) in order to
mitigate the numerical problems caused by degeneracy in the proposal
update scheme.

We first consider as a target density the marginal posterior pdf of
the parameters $\bftheta$ given the observation vector $\textbf{y}$,
i.e., $\pi(\bftheta) = p(\bftheta | \textbf{y})$. As in
\citep{Koblents2013a}, we construct the proposal pdf
$q_\ell(\bftheta)$, $\ell=2,\ldots,L$, as a Gaussian approximation
of the target pdf obtained at the previous iteration $\ell-1$, whose
mean and covariance parameters are selected to match the moments of
the previous sample set. The NPMC algorithm is displayed in Table
\ref{tNPMCalgorithm}. Details and some simple convergence results
can be found in \citep{Koblents2013a}.

\begin{table}[h]
\caption{Nonlinear PMC  targeting $\pi (\bftheta) = p(\bftheta |
\textbf{y})$.} \vspace{-0.3cm} \linethickness{0.3mm} \line(1,0){240}
\vspace{0.1cm} \underline{\textbf{Iteration ($\ell = 1, \ldots,
L$):}}
\begin{enumerate}
\item Draw a set of $M$ samples $\{
\bftheta_\ell^{(i)} \}_{i=1}^M$ from the proposal density
$q_\ell(\bftheta)$:
\begin{itemize}
\item at iteration $\ell=1$, let
$q_1 (\bftheta) = p(\bftheta)$.

\item at iterations $\ell=2,\ldots,L$ the proposal $q_\ell(\bftheta)$ is the Gaussian approximation of $p(\bftheta | \textbf{y})$
obtained at iteration $\ell-1$.
\end{itemize}

\item For $i=1,\ldots,M$, run a SMC scheme with $J$ particles targeting $p(\textbf{x} | \textbf{y},
\bftheta_\ell^{(i)})$ and compute the marginal likelihood estimate
$\hat{p}_\ell^J(\textbf{y} | \bftheta_\ell^{(i)})$.

\item For $i=1,\ldots,M$, compute the unnormalized IWs
\begin{equation*}
w_\ell^{(i)*} \propto \frac{ \hat{p}_\ell^J ( \textbf{y} |
\bftheta_\ell^{(i)} ) p ( \bftheta_\ell^{(i)} ) }{q_\ell (
\bftheta_\ell^{(i)} )}.
\end{equation*}

\item For $i=1,\ldots,M$, compute normalized TIWs, $\bar{w}_\ell^{(i)}$, by \emph{clipping} the original IWs as
\begin{equation*}
\bar{w}_\ell^{(i)*} = \min ( w_\ell^{(i)*}, \mathcal{T}_\ell^{M_T}
), \quad \bar{w}_\ell^{(i)} = \bar{w}_\ell^{(i)*} / \sum_{j=1}^M
\bar{w}_\ell^{(j)*},
\end{equation*}
where the threshold value $\mathcal{T}_\ell^{M_T}$ denotes the
$M_T$-th highest unnormalized IW $w_\ell^{(i)*}$, with $1 < M_T <
M$.

\item Resample to obtain an unweighted set $\{ \tilde{\bftheta}_\ell^{(i)} \}_{i=1}^M$:
for $i,j = 1, \ldots, M$, let $\tilde{\bftheta}_\ell^{(i)} =
\bftheta_\ell^{(j)}$ with probability $\bar{w}_\ell^{(j)}$.

\item Construct a Gaussian approximation $q_{\ell+1} (\bftheta) = \mathcal{N} (\bftheta; \boldsymbol{\mu}_\ell, \boldsymbol{\Sigma}_\ell)$
of the posterior $p(\bftheta | \textbf{y})$, where the mean vector
and covariance matrix are computed as
\begin{equation}
\boldsymbol{\mu}_{\ell} = \frac{1}{M} \sum_{i=1}^M
\tilde{\bftheta}_{\ell}^{(i)} \; \mbox{and} \;
\boldsymbol{\Sigma}_{\ell} = \frac{1}{M} \sum_{i=1}^M (
\tilde{\bftheta}_{\ell}^{(i)} - \boldsymbol{\mu}_{\ell} )(
\tilde{\bftheta}_{\ell}^{(i)} - \boldsymbol{\mu}_{\ell}
)^\top.\label{eqCompMean}
\end{equation}
\end{enumerate}
\label{tNPMCalgorithm} \vspace{-0.2cm} \linethickness{0.3mm}
\line(1,0){240}
\end{table}

Equivalently to the pMCMC algorithm, in the NPMC implementation the
densities $p(\textbf{x}|\textbf{y},\bftheta)$ and $p(\textbf{y} |
\bftheta)$ required in steps 2 and 3 are replaced by their SMC
approximations, which are given in Appendix A. The NPMC method may
also use either exact or approximate samples of the stochastic
process, depending on the computational capabilities.

For the \textit{clipping} procedure performed in step 4 we consider,
at each iteration $\ell$, a permutation $i_1, \ldots, i_M$ of the
indices in $\{1, ..., M \}$ such that $w_\ell^{(i_1)*} \geq \ldots
\geq w_\ell^{(i_M)*}$ and choose a \textit{clipping} parameter $M_T
< M$. We select a threshold value $\mathcal{T}_\ell^M =
w_\ell^{(i_{M_T})*}$ and apply \textit{clipping} to the largest IWs
$w_\ell^{(i_k)*} \geq \mathcal{T}_\ell^M$, $k=1, \ldots, M_T-1$.
This transformation leads to $M_T$ flat TIWs in the region of
interest of $\bftheta$, allowing for a robust update of the
proposal. The performance of the algorithm is robust to the
selection of the \emph{clipping} parameter $M_T$
\citep{Koblents2013a}. For simplicity, step 5 performs multinomial
resampling.

At each iteration of the NPMC algorithm we may construct a discrete
approximation of the posterior pdf $p(\bftheta|\textbf{y})$, based
on the set of samples and TIWs, as
\begin{equation*}
\hat{p}^M_\ell ( d\bftheta | \textbf{y}) = \sum_{i=1}^M
\bar{w}_\ell^{(i)} \delta_{\bftheta_\ell^{(i)}} (d\bftheta).
\end{equation*}

The choice of a Gaussian approximation of the proposal
$q_{\ell+1}(\bftheta)$ in step 6 is arbitrary (and done for
simplicity here). Any other family of pdfs can be used without
modifying the rest of the algorithm
\citep{Koblents2013b,Koblents2013c}.

\subsection{NPMC targeting $p(\bftheta, \textbf{x} | \textbf{y})$}

The NPMC method proposed in \citep{Koblents2013a} may be readily
applied to the approximation of the full joint posterior
$p(\bftheta, \textbf{x} | \textbf{y})$, in an manner equivalent to
the pMCMC algorithm. We consider a sampling mechanism of the form
$q(\bftheta) \hat{p}^J (\textbf{x} | \textbf{y}, \bftheta)$, where
samples $\bftheta^{(i)}$ are again generated from the latest
proposal $q(\bftheta)$ and $\textbf{x}^{(i)}$ are drawn form the SMC
approximation $\hat{p}^J(\textbf{x} | \textbf{y}, \bftheta^{(i)})$
obtained via particle filtering (the iteration index has been
omitted for simplicity). Then, the standard, unnormalized IW
associated to the pair $(\bftheta^{(i)}, \textbf{x}^{(i)})$ is
computed as
\begin{eqnarray*}
w^{(i)*} &=& \frac{\hat{p}^J ( \bftheta^{(i)}, \textbf{x}^{(i)} |
\textbf{y} )}{q ( \bftheta^{(i)} ) \hat{p}^J (\textbf{x}^{(i)} |
\textbf{y}, \bftheta^{(i)})} \propto
\\ && \frac{ \hat{p}^J(\textbf{x}^{(i)}, \textbf{y} |
\bftheta^{(i)} ) p( \bftheta^{(i)})} { q ( \bftheta^{(i)} )
\hat{p}^J(\textbf{x}^{(i)} | \textbf{y}, \bftheta^{(i)}) } \propto
\frac{ \hat{p}^J(\textbf{y} | \bftheta^{(i)} ) p( \bftheta^{(i)})} {
q ( \bftheta^{(i)} ) }
\end{eqnarray*}
and is independent of $\textbf{x}$. This reveals that, when samples
$\textbf{x}_\ell^{(i)}$ are drawn from
$\hat{p}^J(d\textbf{x}|\textbf{y},\bftheta)$ the algorithm yields a
discrete approximation of the posterior distribution of the
unobserved populations $\textbf{x}$ constructed as
\begin{equation*}
\hat{p}^M_\ell ( d\textbf{x} | \textbf{y}) = \sum_{i=1}^M
\bar{w}_\ell^{(i)} \delta_{\textbf{x}_\ell^{(i)}} (d\textbf{x}).
\end{equation*}

Even though the proposed NPMC and the pMCMC require very similar
computations for each pair of samples of $\{ \bftheta,
\textbf{x}\}$, and thus have an equivalent computational cost, the
NPMC has a set of important advantages with respect to its MCMC
counterpart. PMC methods in general can be more easily parallelized,
drastically reducing their execution time. Additionally, they
provide independent sets of samples 
at all iterations, and do not require a burn-in period. On the other
hand, the nonlinearity applied in the NPMC mitigates weight
degeneracy, which is the main problem arising in conventional IS
based methods, dramatically increasing its efficiency in
high-dimensional problems. As a consequence, we claim that the
number of samples (and thus, the computational complexity) required
by the NPMC can be significantly lower than that of pMCMC. Finally,
contrary to pMCMC, which requires a careful choice of the proposal
tuning parameter, the proposed method does not require the precise
fitting of any parameters.

An extensive numerical comparison of pMCMC versus NPMC for the
prokaryotic autoregulatory network is presented in Section
\ref{SKM_example}.

\section{Example: Prokaryotic autoregulatory model}
\label{SKM_example}

In this section, we compare the performance of the pMCMC and the
NPMC methods when applied to the problem of approximating the
posterior distributions of the log-rate parameters $p(\bftheta |
\textbf{y})$ and the populations $p(\textbf{x} | \textbf{y})$ in a
simplified prokaryotic autoregulatory model, given some observed
data $\textbf{y}$. This problem has been introduced in
\citep{Golightly2005}, and further analyzed in
\citep{Golightly2011,Wilkinson2011}. This prokaryotic model is
minimal in terms of the level of details included and offers a
simplistic view of the mechanisms involved in gene autoregulation.
However, it contains many of the interesting features of an
auto-regulatory feedback network and does provide sufficient detail
to capture the network dynamics.

\subsection{Prokaryotic autoregulatory model}
\label{Prok_model}

The prokaryotic autoregulatory model is a SKM that involves $V=5$
chemical species and $K=8$ reaction equations, $r_1, \ldots, r_K$,
given by \citep{Golightly2005}
\begin{equation*}
\begin{array}{ll}
  r_1 : x_{DNA} + x_{P_2} \stackrel{c_1} {\longrightarrow} x_{DNA \cdot P_2}, & \quad r_5 : 2x_{P} \stackrel{c_5} {\longrightarrow} x_{P_2}, \\
  r_2 : x_{DNA \cdot P_2} \stackrel{c_2} {\longrightarrow} x_{DNA} + x_{P_2}, & \quad r_6 : x_{P_2} \stackrel{c_6} {\longrightarrow} 2x_{P}, \\
  r_3 : x_{DNA} \stackrel{c_3} {\longrightarrow} x_{DNA} + x_{RNA}, & \quad r_7 : x_{RNA} \stackrel{c_7} {\longrightarrow} 0, \\
  r_4 : x_{RNA} \stackrel{c_4} {\longrightarrow} x_{RNA} + x_{P}, & \quad r_8 : x_{P} \stackrel{c_8} {\longrightarrow} 0.
\end{array}
\end{equation*}

We construct the $V$-dimensional vector containing the population of
each species at time instant $t$ as $\textbf{x}(t) = [x_{RNA}(t),
x_P(t), x_{P_2}(t), x_{DNA \cdot P_2}(t), x_{DNA}(t)]^\top$. Thus,
we obtain a stoichiometry matrix of the form
\begin{equation*}
\textbf{S} = \left(
      \begin{array}{rrrrrrrr}
        0 & 0 & 1 & 0 & 0 & 0 & -1 & 0 \\
        0 & 0 & 0 & 1 & -2 & 2 & 0 & -1 \\
        -1 & 1 & 0 & 0 & 1 & -1 & 0 & 0 \\
        1 & -1 & 0 & 0 & 0 & 0 & 0 & 0 \\
        -1 & 1 & 0 & 0 & 0 & 0 & 0 & 0 \\
      \end{array}
    \right)
\end{equation*}
and the hazard vector is given by
\begin{eqnarray}
\label{eq_hazard} \textbf{h}(t) = &[& c_1 x_{DNA} x_{P_2}, c_2
x_{DNA \cdot P_2}, c_3
x_{DNA}, c_4 x_{RNA}, \nonumber \\
&& c_5 \frac{x_P(x_P-1)}{2}, c_6 x_{P_2}, c_7 x_{RNA}, c_8 x_{P}
]^\top,
\end{eqnarray}
where the time dependance of the population of each species is
omitted for notational simplicity.

This model involves a conservation law given by the relation $x_{DNA
\cdot P_2} + x_{DNA} = C$, where $C$ is the number of copies of this
gene in the genome. We could use this relation to remove $x_{DNA
\cdot P_2}$ from the model, replacing any occurrences of the latter
in the hazard function with $C - x_{DNA}$, but in this paper we
abide by the notation in equation (\ref{eq_hazard}). Further details
of this model can be found in \citep{Wilkinson2011}.

\subsection{Simulation setup}
\label{Num_results}

We have selected most of the simulation parameters following
\citep{Golightly2011}. The true vector of rate parameters which we
aim to estimate has been set to
\begin{equation*}
\textbf{c} = [0.1, 0.7, 0.35, 0.2, 0.1, 0.9, 0.3, 0.1]^\top,
\end{equation*}
which yields log-transformed rate parameters
\begin{equation*}
\bftheta = -[ 2.30, 0.36, 1.05, 1.61, 2.30, 0.10, 1.20, 2.30 ]^\top.
\end{equation*}
The initial populations and the conservation constant have been set
to $\textbf{x}_0 = [x_1(0), \ldots, x_V(0)]^\top = [ 8, 8, 8, 5,5
]^\top$ and $C = 10$, respectively. The time discretization period
is $\Delta = 1$ and the Gaussian noise variance is $\sigma^2 = 4$
(assumed to be known). In all the simulations in this paper we have
performed exact sampling from the stochastic model with the
Gillespie algorithm to obtain the likelihood approximation via
particle filtering. The number of particles for the SMC
approximation $\hat{p}^J( \textbf{x} | \bftheta, \textbf{y})$, has
been set to $J=100$ for all the simulations.

Independent uniform priors $\mathcal{U}(\theta_k; -7,2)$ are taken
for each $\theta_k = \log(c_k)$. Opposite to \citep{Golightly2011},
the initial populations $\textbf{x}_0$ are assumed unknown for the
inference algorithm and we consider independent Poisson priors
$p(x_v(0)) = \mathcal{P}(x_v(0); \lambda_v )$, with $\lambda_v$
parameters set to the true initial populations, that is, $\lambda_v
= x_v(0)$, $v = 1,\ldots,V$.

We consider two different observation scenarios. In the complete
observation (CO) scenario we assume that all species $x_v$, $v = 1,
\ldots, V$, are observed at regular time intervals of length
$\Delta$ and corrupted by Gaussian noise. Thus, the observation
matrix is of the form $\textbf{M} = \textbf{I}_V$ and the
observations are given by
\begin{equation*}
\textbf{y}_n = \textbf{x}_n + \textbf{w}_n, \; n = 1,\ldots,N.
\end{equation*}
In the CO case the complete vector of observations $\textbf{y} =
[\textbf{y}_1^\top, \ldots, \textbf{y}_N^\top]^\top$ has dimension
$VN \times 1$.

In the partial observation scenario (PO) only a linear combination
of the proteins $x_{P} + 2 x_{P_2}$ is observed, also contaminated
by Gaussian noise, i.e., the observation matrix is given by
$\textbf{M} = [0,1,2,0,0]$ (with dimension $1 \times V$) and the
observations are generated as
\begin{equation*}
y_n = x_{2,n} + 2 x_{3,n} + w_n, \; \textrm{where} \; w_n \sim
\mathcal{N}_1(w_n; 0, \sigma^2).
\end{equation*}
In the PO case, a vector of scalar observations with dimension $N
\times 1$ is constructed as $\textbf{y} = [ y_1, \ldots,
y_{N}]^\top$.

\subsection{Performance evaluation}

To evaluate the performance of the pMCMC and the NPMC methods we
compute, in all the simulation runs, the mean square error (MSE)
attained by the sample set that approximates the marginal posterior
of $\bftheta$, generated by both schemes.

For the pMCMC method, we compute the MSE of each parameter
$\theta_k$ based on the $M$-size final output (after removing the
burn-in period and thinning), as
\begin{equation*}
MSE_k = \frac{1}{M} \sum_{i=1}^M ( \theta_k^{(i)} - \theta_k )^2$,
$k \in \{ 1, ..., K \}.
\end{equation*}

For the NPMC, we compute the MSE associated to each parameter
$\theta_k$, $k=1,\ldots,K$, based on the unweighted sample set at
the $\ell$-th iteration $\{ \tilde{\bftheta}_\ell^{(i)} \}_{i=1}^M$,
$\ell=1,\ldots,L$, as
\begin{equation*}
MSE_{\ell,k} = \frac{1}{M} \sum_{i=1}^M (
\tilde{\theta}_{\ell,k}^{(i)} - \theta_k )^2 = (\mu_{\ell,k} -
\theta_k)^2 + \sigma_{\ell,k}^2,
\end{equation*}
where $\mu_{\ell,k}$ is the $k$-th component of the mean vector
$\boldsymbol{\mu}_\ell$ and the variance term $\sigma_{\ell,k}^2$ is
the $(k,k)$ component of matrix $\boldsymbol{\Sigma}_\ell$.

However, the MSE cannot be computed in real problems, where the true
parameters $\theta_k$ are unknown. To monitor the stability and the
efficiency of the two sampling schemes based on the generated sample
alone, we resort to the so called normalized effective sample size
(NESS), which is often defined differently for MCMC and IS schemes
\citep{Robert04}.

In the MCMC literature, the NESS gives the relative size of an
i.i.d. (independent and identically distributed) sample with the
same variance as the current sample and thus indicates the loss in
efficiency due to the use of a Markov chain \citep{Robert04}. For
pMCMC we compute the NESS from the final chain (after removing the
burn-in period and thinning) as
\begin{equation*}
M^{neff} = \frac{1}{1 + 2\sum_{j=1}^\infty \hat{\rho}(j)},
\end{equation*}
where $\hat{\rho}(j) = \textrm{corr}(\bftheta^{(0)},
\bftheta^{(j)})$ is the average autocorrelation function (ACF) at
lag $j$. For the computation of the NESS, we truncate $j$ when
$\hat{\rho}(j) < 0.1$.

For IS methods, the NESS may be interpreted as the relative size of
a sample generated from the target distribution with the same
variance as the current sample. Even when high values of the NESS do
not guarantee a low approximation error, the NESS is often used as
an indicator of the numerical stability of the algorithm
\citep{Doucet00}. It cannot be evaluated exactly but we may compute
an approximation of the NESS at each iteration of the NPMC scheme
based on the set of TIWs as
\begin{equation*}
M_\ell^{neff} = \frac{1}{M \sum_{i=1}^M (\bar{w}_\ell^{(i)})^2},
\quad \ell=1,\ldots,L.
\end{equation*}

\begin{figure*}[htp]
\centering
\includegraphics[width=0.45\textwidth]{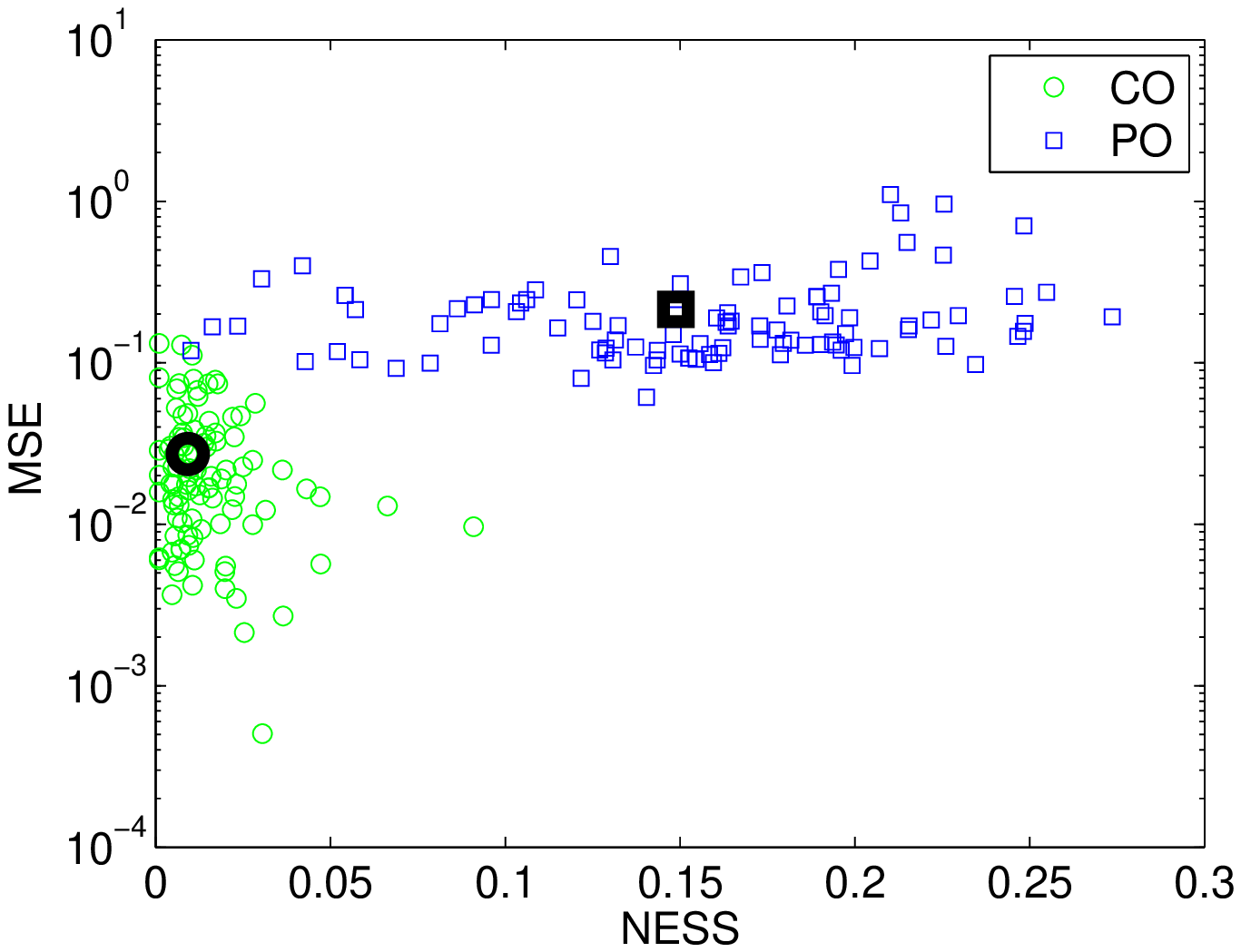}
\includegraphics[width=0.45\textwidth]{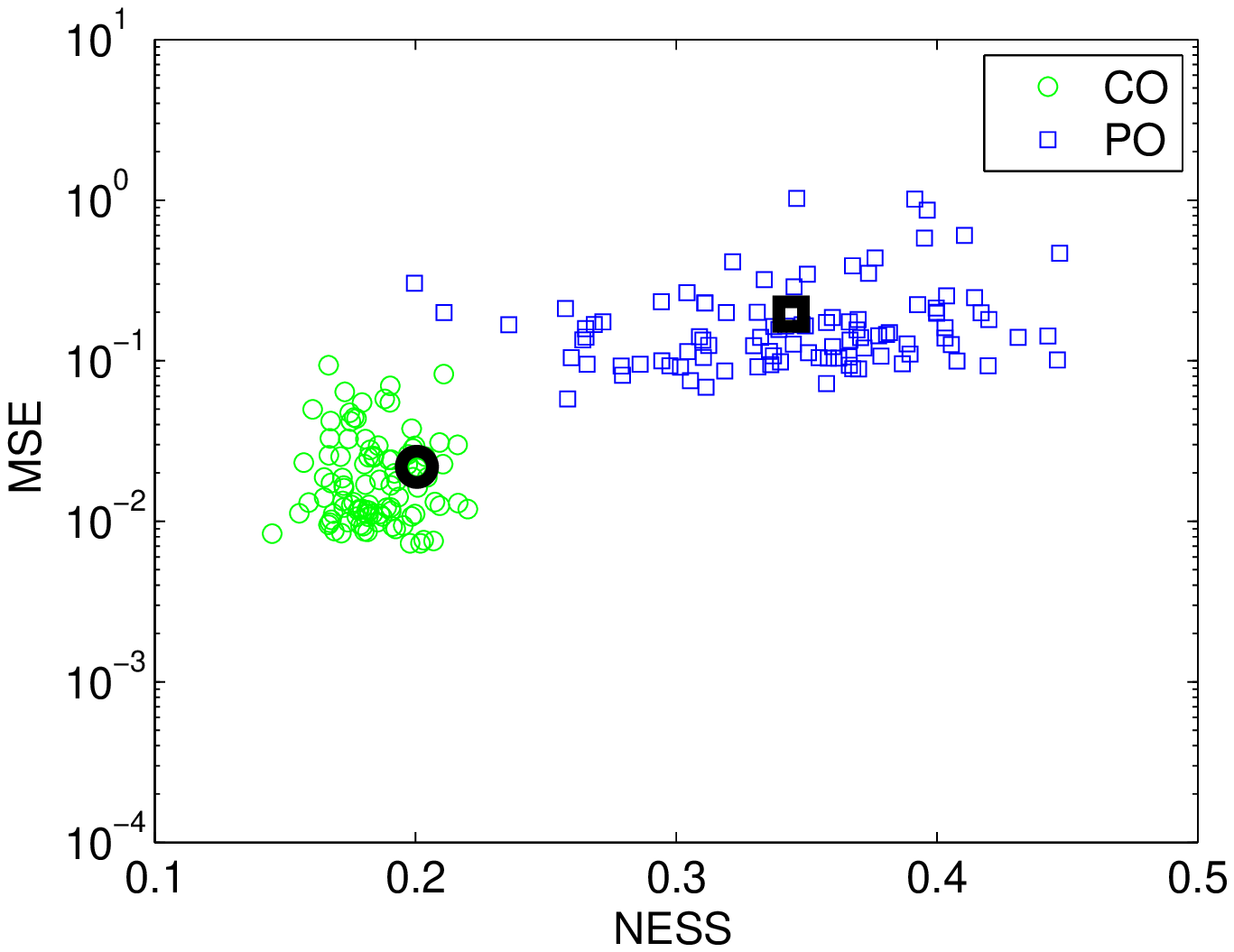}
\caption{Performance of the pMCMC (\textit{left}) and the NPMC
(\textit{right}) methods for the estimation of a unique rate
parameter $\theta_1$: MSE (in logarithmic scale) obtained from the
final output versus the NESS for each simulation run in the CO and
the PO scenario. The big circles and squares represent simulation
runs with a final mean MSE close to the global average}
\label{fig_scatter_1param}
\end{figure*}

\begin{figure*}[htp]
\centering
\includegraphics[width=0.45\textwidth]{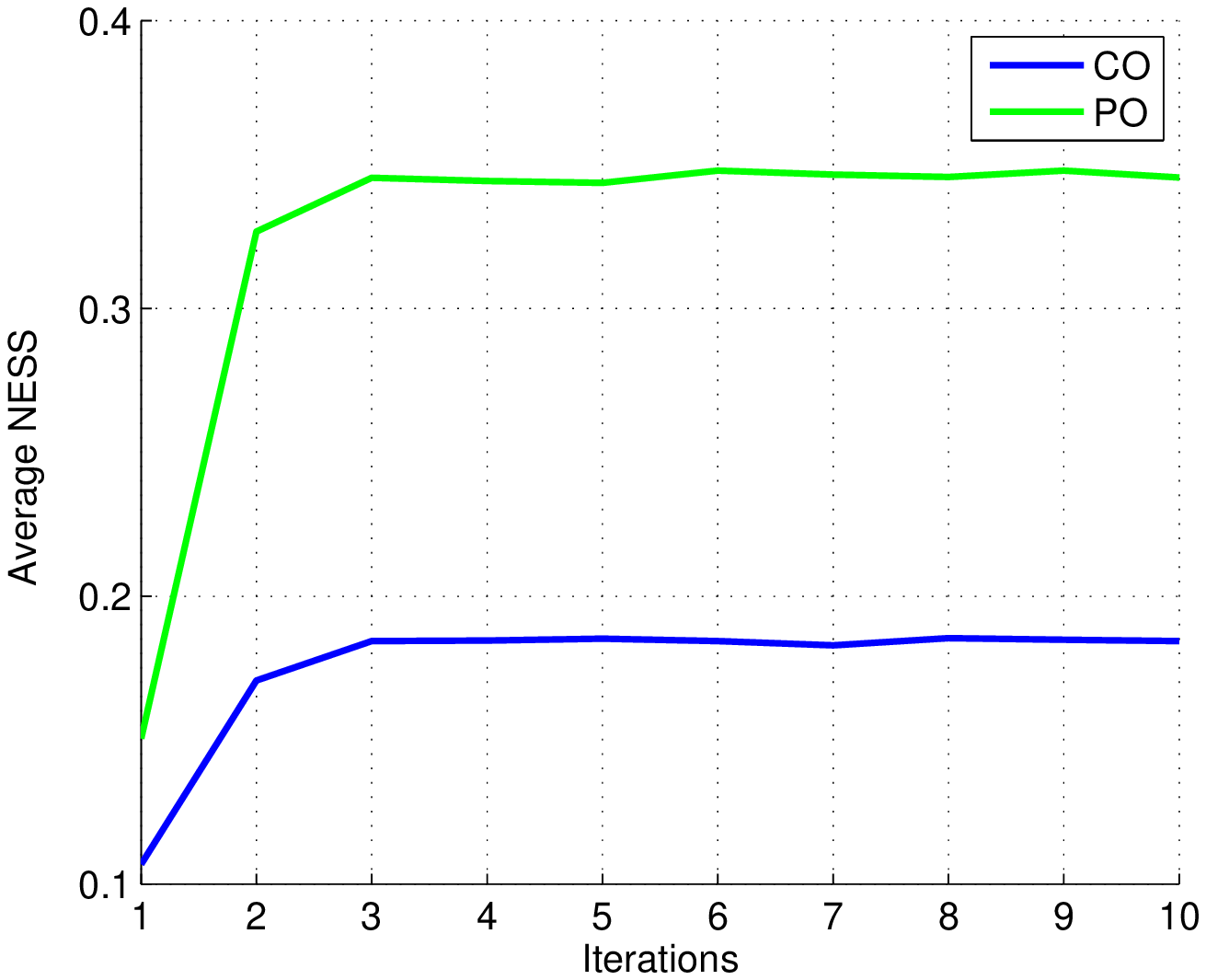}
\includegraphics[width=0.45\textwidth]{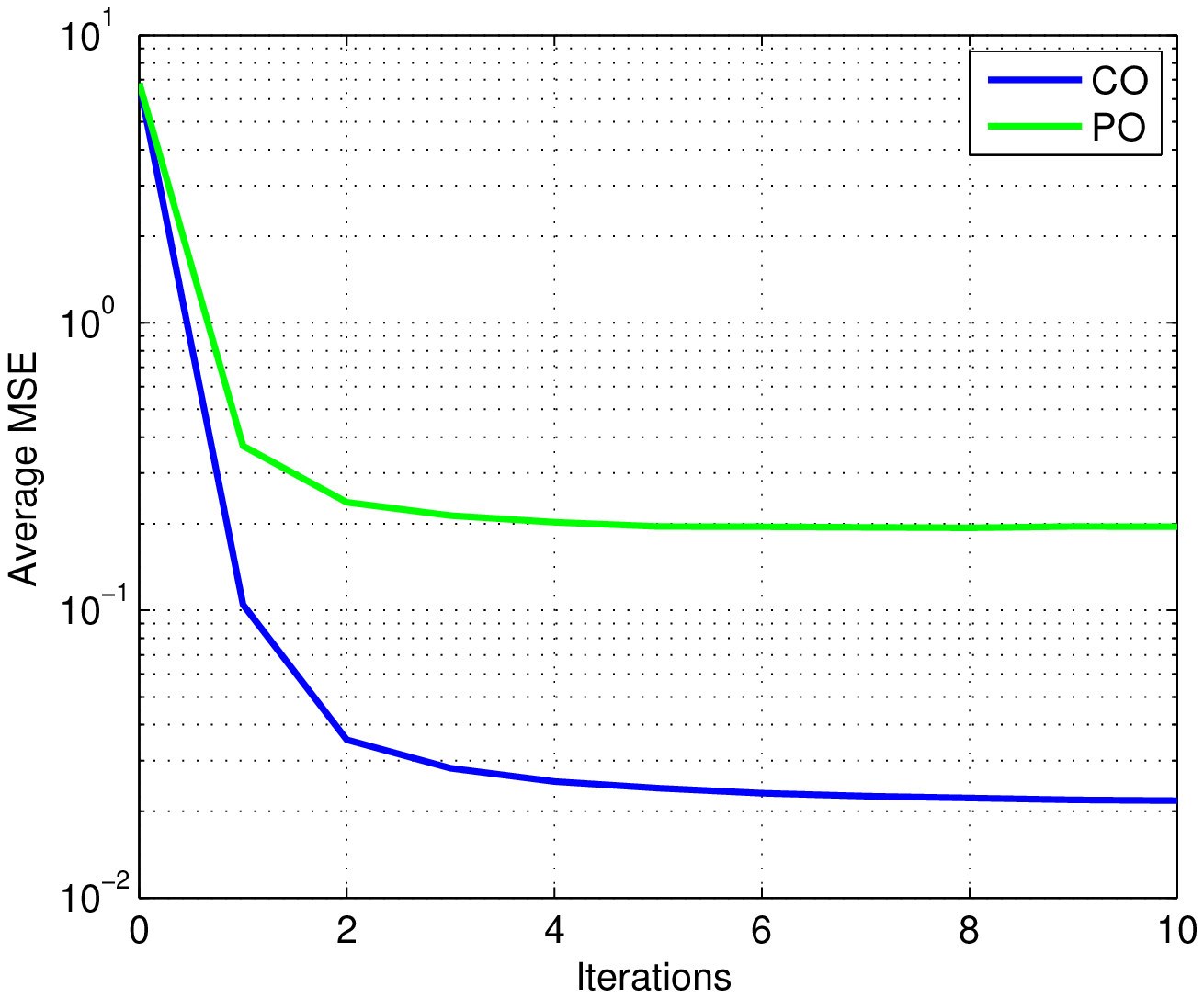}
\caption{Evolution along the iterations of the NPMC algorithm of the
average NESS (\textit{left}) and MSE (\textit{right}) in the CO and
PO scenarios, estimating a single parameter $\theta_1$.}
\label{fig_1param_NPMC}
\end{figure*}

\subsection{Simulation results}

We consider two simulation scenarios in which a different number of
parameters is estimated.

\subsubsection{Estimation of a single rate parameter $\theta_1$}
\label{sims_1param}

In this section we present numerical results regarding the
approximation of the posterior distribution $p(\theta_1, \textbf{x}
| \bftheta_{\backslash 1}, \textbf{y})$ of a single rate parameter
$\theta_1 = \log c_1$ and the populations $\textbf{x}$, when the
rest of parameters $\bftheta_{\backslash 1} = [\theta_2, \ldots,
\theta_K]^\top$, are assumed to be known.

We compare the pMCMC and the NPMC methods in this simple scenario
in order to illustrate the optimal performance of both schemes, in
the CO and PO scenarios. This simulations show the degradation of
the approximations when the amount of observations reduces.

We have performed $P=100$ independent simulation runs of the pMCMC
and the NPMC schemes in the CO and the PO scenarios, with different
(independent) population and observation vectors in each simulation.
Both in the CO and the PO cases, the same true population
trajectories $\textbf{x}^{(p)}$, $p=1, \ldots, P$, were used, but
the observations in the CO scenario, $\textbf{y}_{CO}^{(p)}$, and in
the PO scenario, $\textbf{y}_{PO}^{(p)}$, differ. The number of
observation times has been set to $N=100$.

As a proposal pdf $q(\bftheta^\star | \bftheta)$ in the pMCMC scheme
we consider a Gaussian random walk update with variance $\gamma^2 =
1$, which to the best results in the simulations. A total number of
$I = 10^4$ iterations has been run in each simulation. A final
sample of size $M=10^3$ has been obtained from each Markov chain by
discarding a burn-in period of $10^3$ samples and thinning the
output by a factor of 9.

In the NPMC scheme, the number of iterations has been set to $L=10$,
the number of samples per iteration is $M=10^3$ and the
\textit{clipping} parameter is $M_T=100$. In this way, the
computational effort of the two methods is approximately the same,
as they both generate $10^4$ samples in the space of $\bftheta$.

In Figure \ref{fig_scatter_1param} the final MSE obtained by the
pMCMC (\textit{left}) and the NPMC (\textit{right}) algorithms for
each simulation run is depicted versus the final NESS, in the CO and
the PO scenarios. Note that the NESS is computed differently for
pMCMC and NPMC. It can be observed that both algorithms perform
similarly in this case, with an equivalent computational cost. Both
algorithms attain on average lower MSE values in the CO scenario, as
expected. However, the NESS also takes lower values in the CO case,
which indicates a worse mixing of the Markov chains in the pMCMC
algorithm and also higher degeneracy in the NPMC algorithm.

\begin{figure}[t]
\centering
\includegraphics[width=0.49\textwidth]{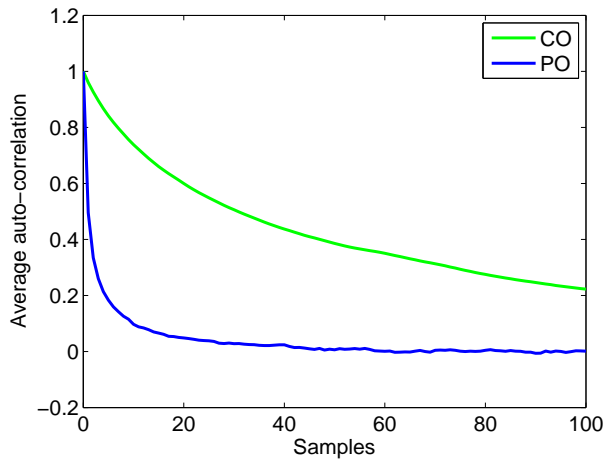}
\caption{Average ACF based on the final sample of size $M=10^3$ of
the pMCMC scheme in the CO and the PO scenarios, averaged over
$P=100$ simulation runs} \label{fig3_1param_pMCMC}
\end{figure}

In Figure \ref{fig_1param_NPMC} the evolution of the MSE
(\textit{right}) and the NESS (\textit{left}) along the iterations
of the NPMC algorithm is represented, for the CO and the PO
scenarios. It can be observed that both measures attain a steady
value by the 5-th iteration, both in the CO and the PO case, which
suggest that actually less iterations are sufficient for this
problem. Again, we observe that in the CO scenario both the NESS and
the MSE reach lower values.

Figure \ref{fig3_1param_pMCMC} plots the average ACF of the final
pMCMC sample, after removing the burn-in period and thinning the
Markov chain by a factor of 9. Particularly high correlations are
present in the CO case, leading to a poor NESS. Related to the ACF,
the average sample acceptance probability in the pMCMC scheme in the
PO scenario is 0.091, while in the CO scenario it is only 0.0034.
Which means that 910 samples are accepted out of $I=10^4$ in the CO
case and only 34 in the CO case.

\begin{figure}[t]
\centering
\includegraphics[width=0.49\textwidth]{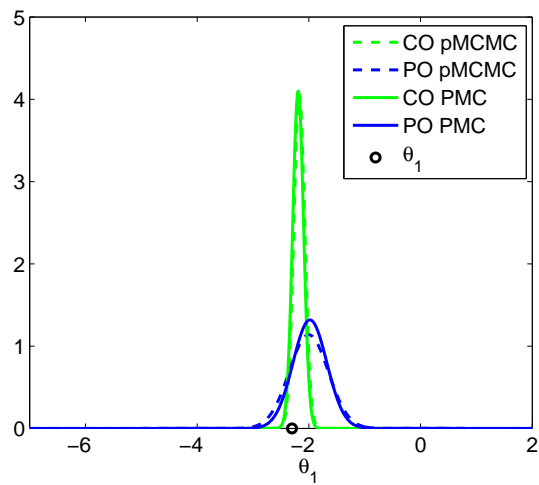}
\caption{Marginal posterior pdf estimates $\hat{p}(\theta_1, |
\bftheta_{\backslash 1}, \textbf{y})$ of an average simulation run,
for pMCMC and NPMC in the CO and PO scenarios. The true value
$\theta_1$ is also shown} \label{fig_dens_1param}
\end{figure}

In Figure \ref{fig_dens_1param} the final pdf estimates
$\hat{p}(\theta_1 | \bftheta_{\backslash 1}, \textbf{y})$ of the
average simulation runs represented as big circles and crosses in
Figure \ref{fig_scatter_1param} are represented in the CO and the PO
scenario, for the pMCMC and the NPMC schemes. For the pMCMC method
we have built a Gaussian approximation of the posterior density
$p(\theta_1 | \bftheta_{\backslash 1}, \textbf{y})$ based on the
final MCMC sample $\{ \theta_1^{(i)} \}_{i=1}^M$. For the NPMC
method, this approximation corresponds to the proposal pdf for the
next iteration $L+1$, i.e., $\hat{p}(\theta_1 | \bftheta_{\backslash
1}, \textbf{y}) = q_{L+1} (\theta_1) = \mathcal{N}(\theta_1;
\mu_{L,1}, \sigma_{L,1}^2)$, where the mean and variance terms
$\mu_{L,1}$ and $\sigma_{L,1}^2$ are computed as in Eq.
(\ref{eqCompMean}). It can be observed in Figure
\ref{fig_dens_1param} that very similar results are obtained by both
algorithms in this scenario. The final MSE values obtained by the
pMCMC and the NPMC methods, averaged over $P=100$ simulation runs,
are shown in Table \ref{table_1param}, together with the MSE
corresponding to the prior distribution.

\begin{table} \centering
\caption{Final mean and standard deviation (std) values of the MSE
for $\theta_1$ in the CO and PO scenarios, for pMCMC and NPMC. The
prior values are included for comparison} \label{table_1param}
\begin{tabular}{llll}
\hline\noalign{\smallskip}
  & & mean MSE & std MSE \\
\noalign{\smallskip}\hline\noalign{\smallskip}
  Prior & & 6.789 & 0 \\
\noalign{\smallskip}\hline\noalign{\smallskip}
\multirow{2}{*}{PO} & pMCMC & 0.215 & 0.171 \\
  & NPMC & 0.195 & 0.170 \\
\multirow{2}{*}{CO} & pMCMC & 0.027 & 0.026\\
  & NPMC & 0.022 & 0.016 \\
\noalign{\smallskip}\hline\noalign{\smallskip}
\end{tabular}
\end{table}

\begin{figure*}[t]
\centering
\includegraphics[width=0.49\textwidth]{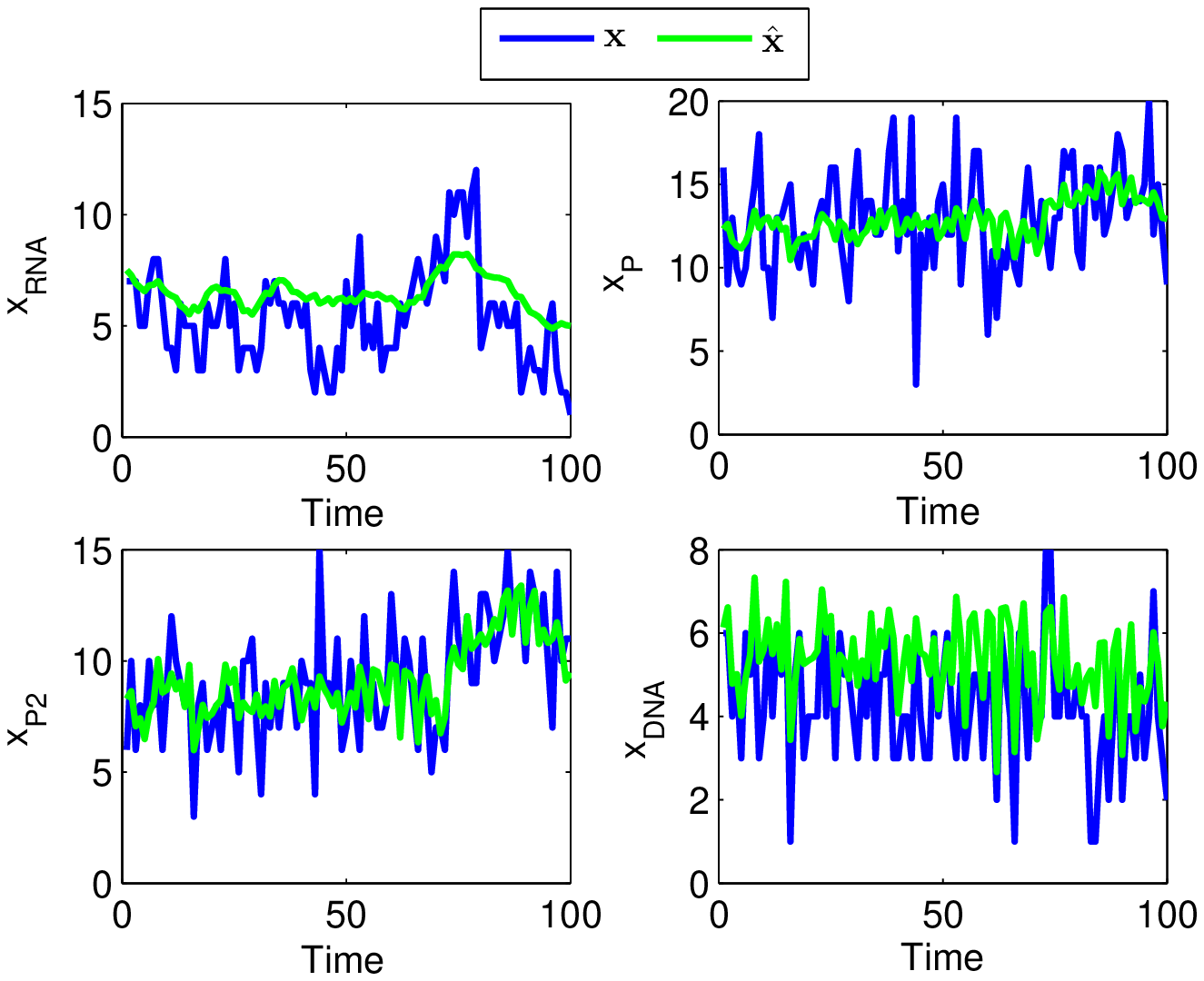}
\includegraphics[width=0.49\textwidth]{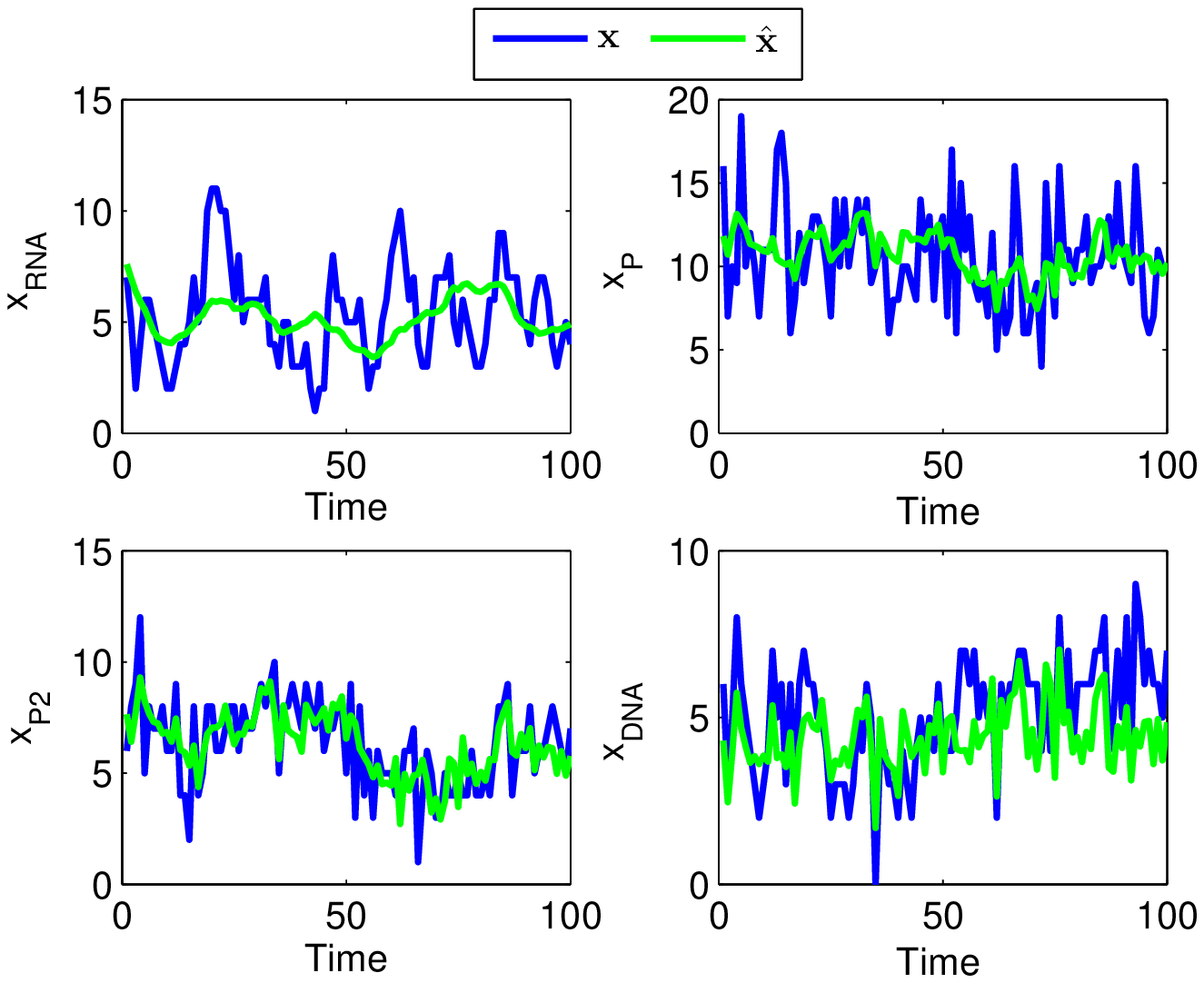}
\caption{Posterior mean, $\hat{\textbf{x}} = E_{\hat{p} (\textbf{x}
| \textbf{y})} [\textbf{x}]$, of the populations obtained in a
single simulation run of pMCMC (\textit{left}) and NPMC
(\textit{right}) in the PO scenario (only a linear combination of
the proteins is observed, corrupted by noise)}
\label{fig_x_est_1param}
\end{figure*}

Figure \ref{fig_x_est_1param} depicts the posterior mean of the
populations, $\hat{\textbf{x}} = E_{\hat{p} (\textbf{x} |
\textbf{y})} [\textbf{x}]$, obtained with pMCMC (\textit{left}) as
$\hat{\textbf{x}} = \frac{1}{M} \sum_{i=1}^M \textbf{x}^{(i)}$ and
with NPMC (\textit{right}) as $\hat{\textbf{x}} = \sum_{i=1}^M
\bar{w}_L^{(i)} \textbf{x}_L^{(i)}$ in the PO scenario. The results
correspond to the particular simulation runs (different for pMCMC
and NPMC) identified with big squares in Figure
\ref{fig_scatter_1param} and whose posterior approximations,
$\hat{p}(\theta_1 | \bftheta_{\backslash 1}, \textbf{y})$, are shown
in Figure \ref{fig_dens_1param}. It can be observed that, in the PO
scenario, the tendency of the population of all the species is
reasonably identified, even though only a linear combination of the
proteins is observed. In the CO scenario the populations of all
species are accurately estimated and are not shown for conciseness.
Note that the populations of all species are very low, which
suggests that the diffusion approximation may perform poorly in this
scenario.

The results presented in this section reveal a very similar
performance of the two methods in this simple scenario. Also in
terms of computational complexity pMCMC and NPMC perform very
similarly. The execution time per $10^3$ samples (one NPMC iteration
and $10^3$ pMCMC iterations) for the pMCMC scheme is 312 seconds,
while for NPMC it is 325 seconds, both in the CO and in the PO
cases, on a 3-GHz Intel Core 2 Duo CPU, with 2 GB of RAM. The
stochastic forward simulation of the prokaryotic model with the
Gillespie algorithm has been implemented in C, and the rest of the
code in Matlab R2007b.

However, the pMCMC method provides a set of highly correlated
samples, specially in the CO scenario, and requires the setting of
the proposal variance $\gamma^2$ as well as the burn-in period
length and the thinning parameter, which may not be straightforward
and determines the performance of the algorithm. On the contrary,
the NPMC scheme provides uncorrelated sets of samples 
at each iteration, and does not require the precise fitting of any
parameters. Additionally, the computer simulations suggest that the
convergence of the NPMC algorithm may be assessed observing the
evolution of the NESS, which usually reaches a steady value
simultaneously with the MSE.

\subsubsection{Estimation of all the parameters $\theta_k$,
$k=1,\ldots,K$}

In this section we present simulation results to evaluate the
performance of the pMCMC and the NPMC schemes in the approximation
of the posterior distribution of the rate parameters and the
populations of all species, $p(\bftheta, \textbf{x} | \textbf{y})$,
assuming that all the rate parameters are unknown, again in the CO
and the PO scenarios.

In this case, $N=200$ observation times are assumed for all the
simulations. Again, $P=100$ independent simulation runs of each
algorithm have been performed. The NPMC scheme has been run for
$L=15$ iterations, with $M=10^3$ samples per iteration and
\textit{clipping} parameter $M_T = 100$. The pMCMC scheme has been
run with $I = 15 \times 10^3$ iterations in each simulation run, a
burn-in period of $10^3$ iterations and thinning the output by a
factor of $14$. With this setup the computational effort is
approximately the same in the two schemes.

\begin{figure*}[htp]
\centering 
\includegraphics[width=0.45\textwidth]{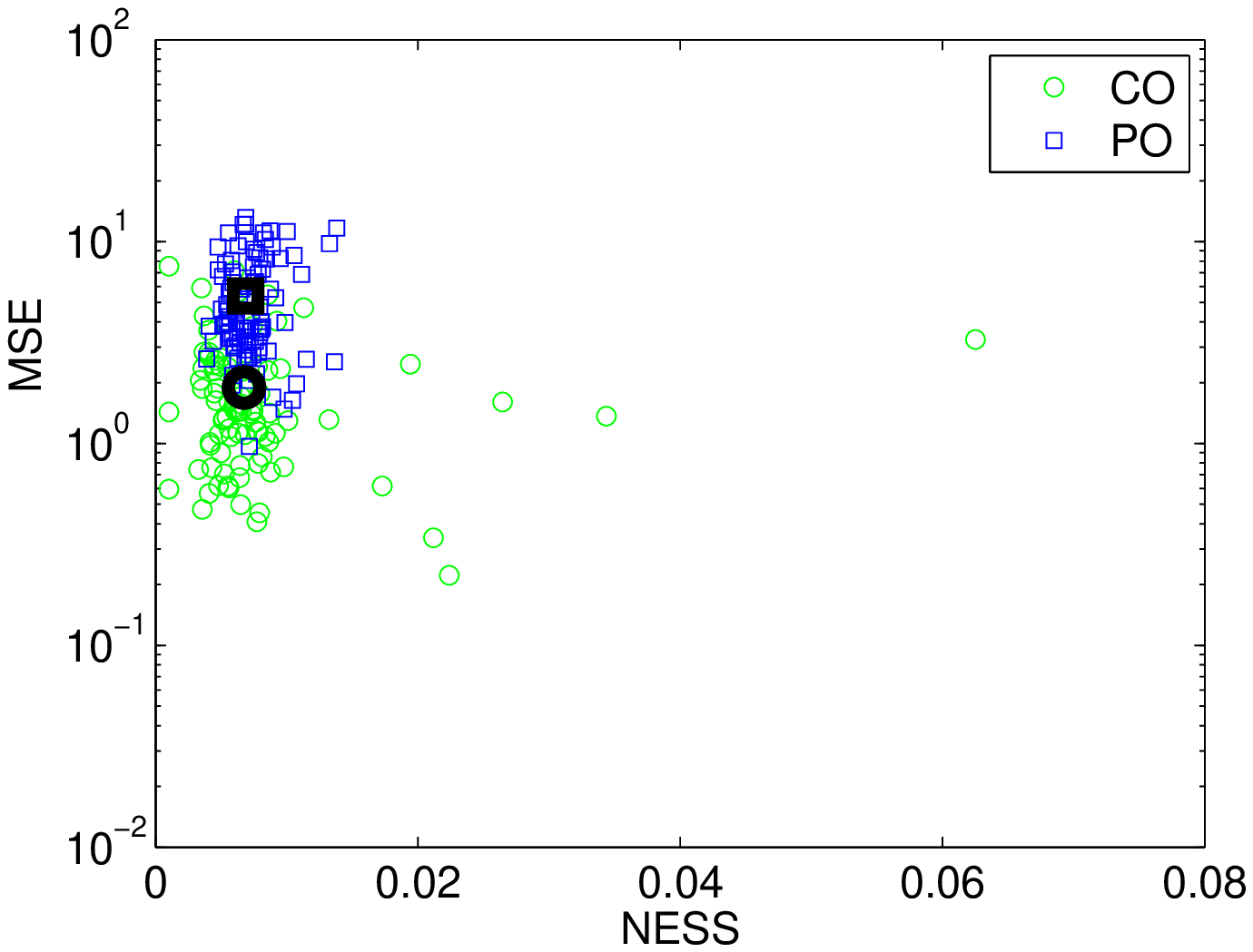}
\includegraphics[width=0.45\textwidth]{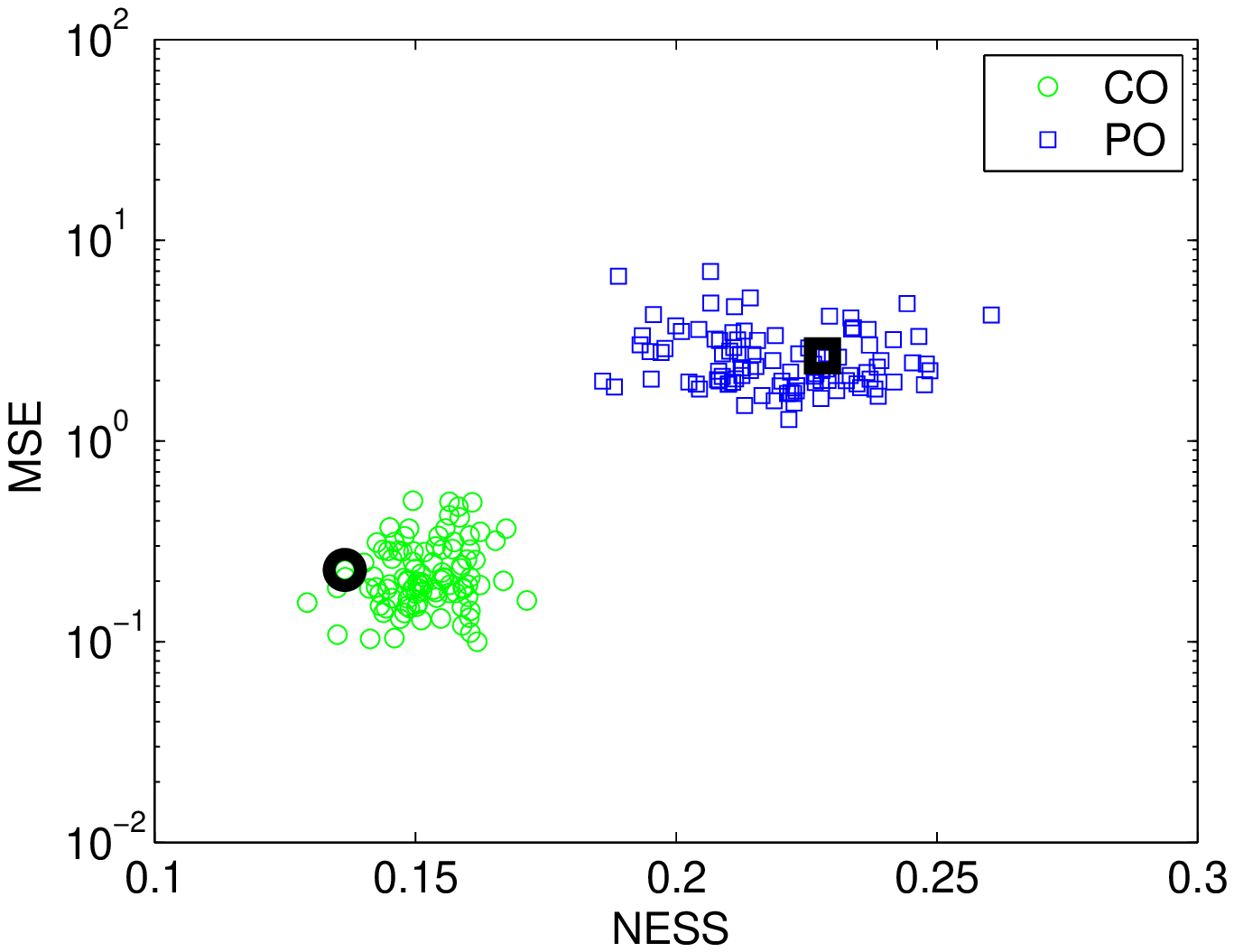}
\caption{Performance of the pMCMC (\textit{left}) and the NPMC
(\textit{right}) methods for the estimation of the whole set of rate
parameters $\bftheta$: MSE (in logarithmic scale) versus the final
NESS, for each simulation run in the CO and the PO scenario. The big
circles and squares represent simulation runs with a final mean MSE
close to the global average} \label{fig_scatter_8params}

\vspace{0.5cm}

\centering 
\includegraphics[width=0.45\textwidth]{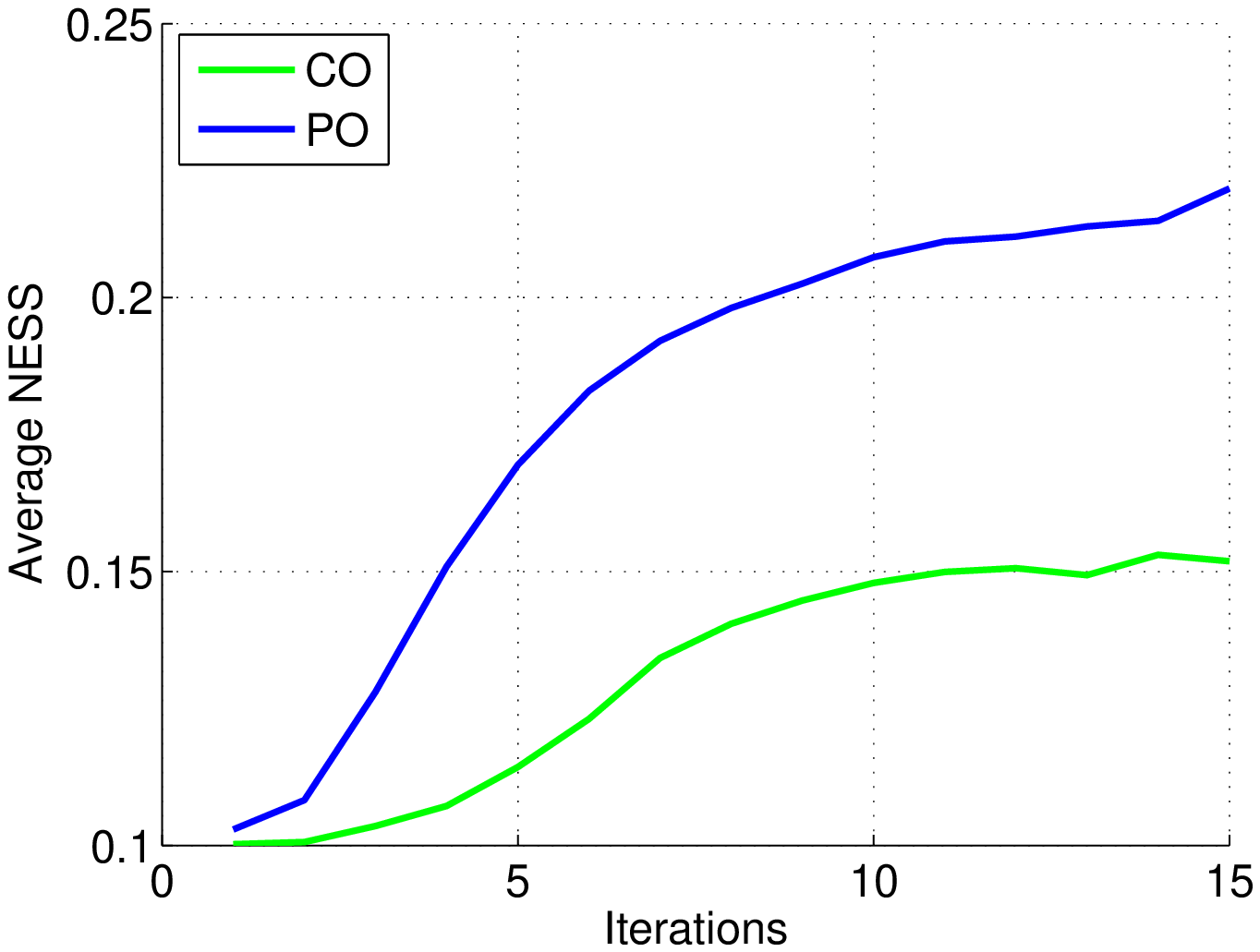}
\includegraphics[width=0.45\textwidth]{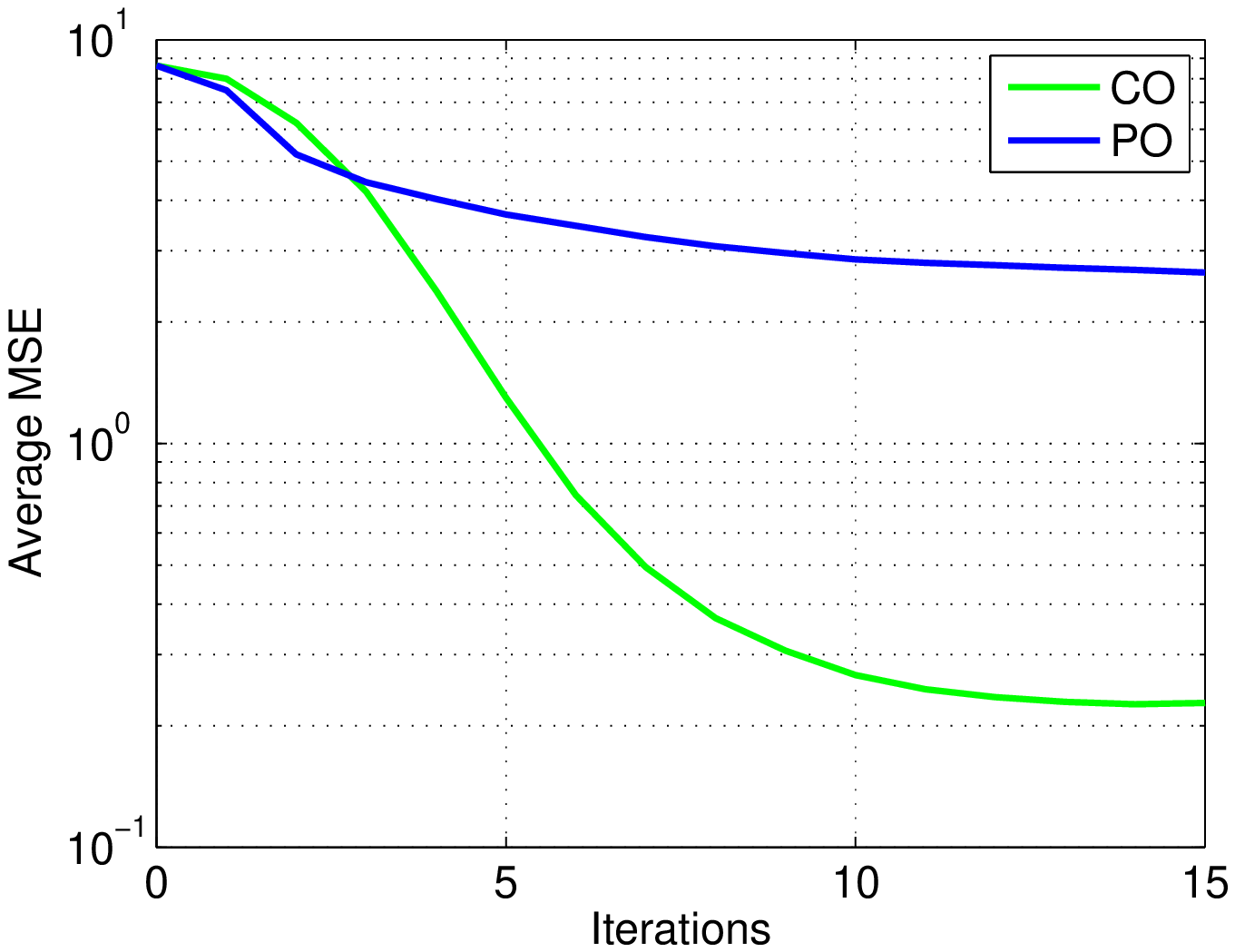}
\caption{Evolution along the NPMC iterations of the average NESS
(\textit{left}) and MSE (\textit{right}) in the CO and the PO
scenario.} \label{fig_8params_NPMC}

\vspace{0.5cm}

\centering
\includegraphics[width=0.45\textwidth]{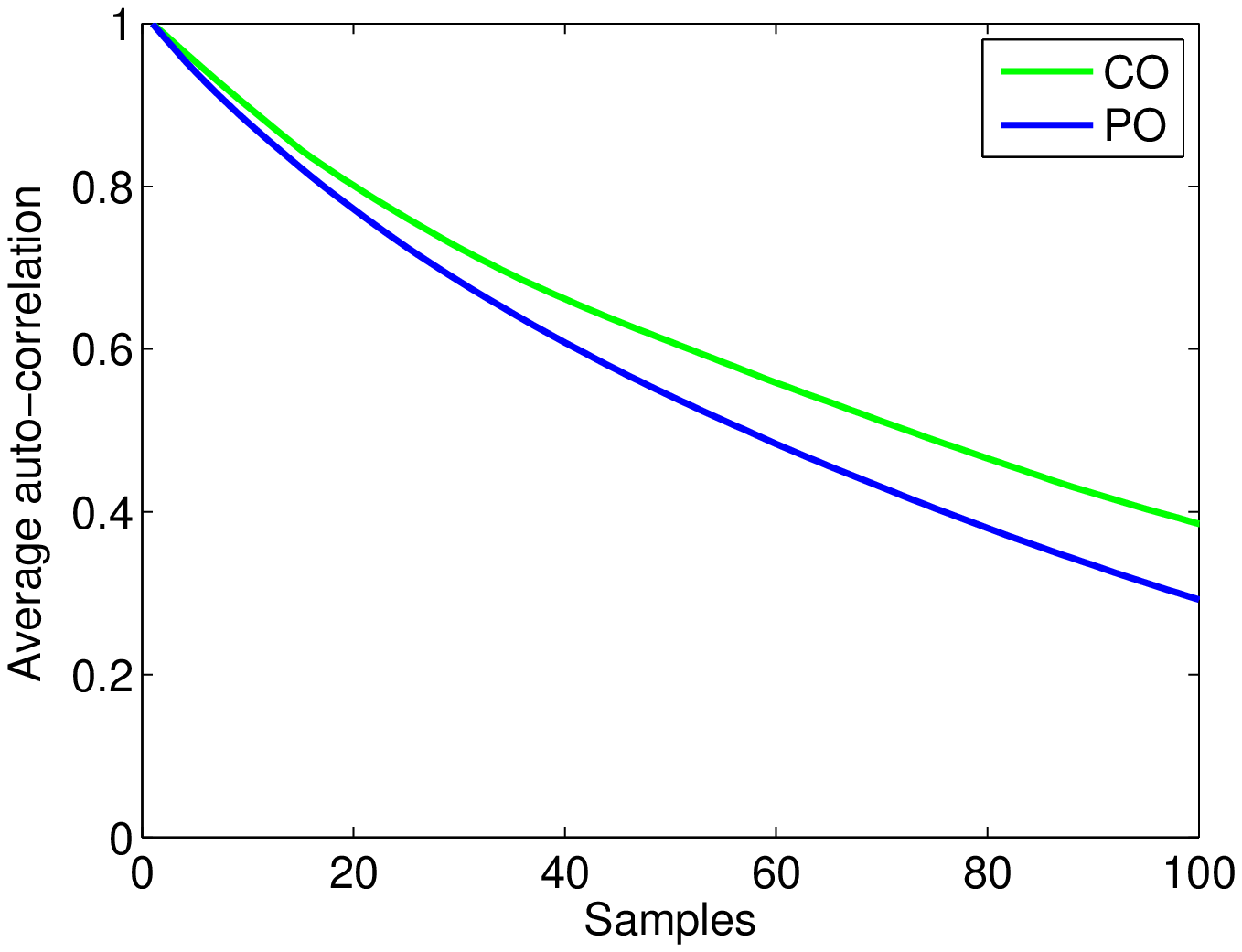}
\includegraphics[width=0.45\textwidth]{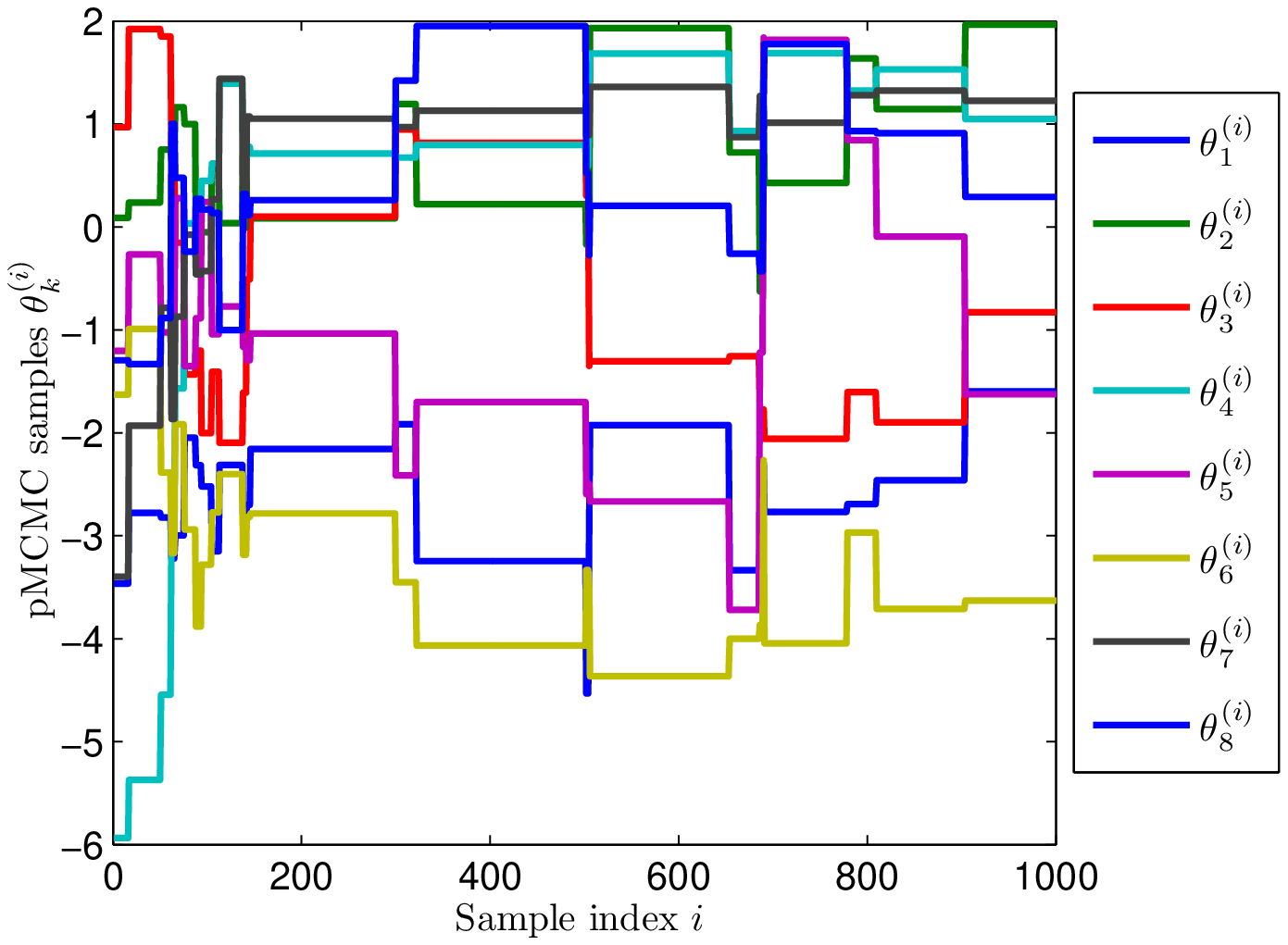}
\caption{\textit{Left}: Auto-correlations based on the final sample
of size $10^3$ of the pMCMC scheme in the CO and the PO scenarios,
averaged over $P=100$ simulation runs. \textit{Right}: Markov chain
provided by the pMCMC method in the PO scenario, corresponding to
the average simulation run depicted with a big square in Figure
\ref{fig_scatter_8params} (\textit{left}).}
\label{fig3_8params_pMCMC}
\end{figure*}

In Figure \ref{fig_scatter_8params} the MSE (in logarithmic scale),
averaged over the parameters $\theta_k$, attained by the pMCMC
(\textit{left}) and the NPMC (\textit{right}) algorithms is
represented versus the NESS, in the CO and PO scenarios. Simulation
runs which attained a final MSE close to the global average value
are indicated with big circles (CO) and squares (PO) on both plots.
It can be observed that the pMCMC method performs similarly in both
scenarios, in terms of MSE and NESS, yielding poor results in both
cases. On the contrary, the NPMC method provides significantly
better $MSE$ results in the CO scenario, where a larger amount of
information is available. The NPMC method does not present
degradation due to the high degeneracy occurring in the CO scenario.

\begin{figure*}[htp]
\hspace{-1cm}
\includegraphics[width=1.1\textwidth]{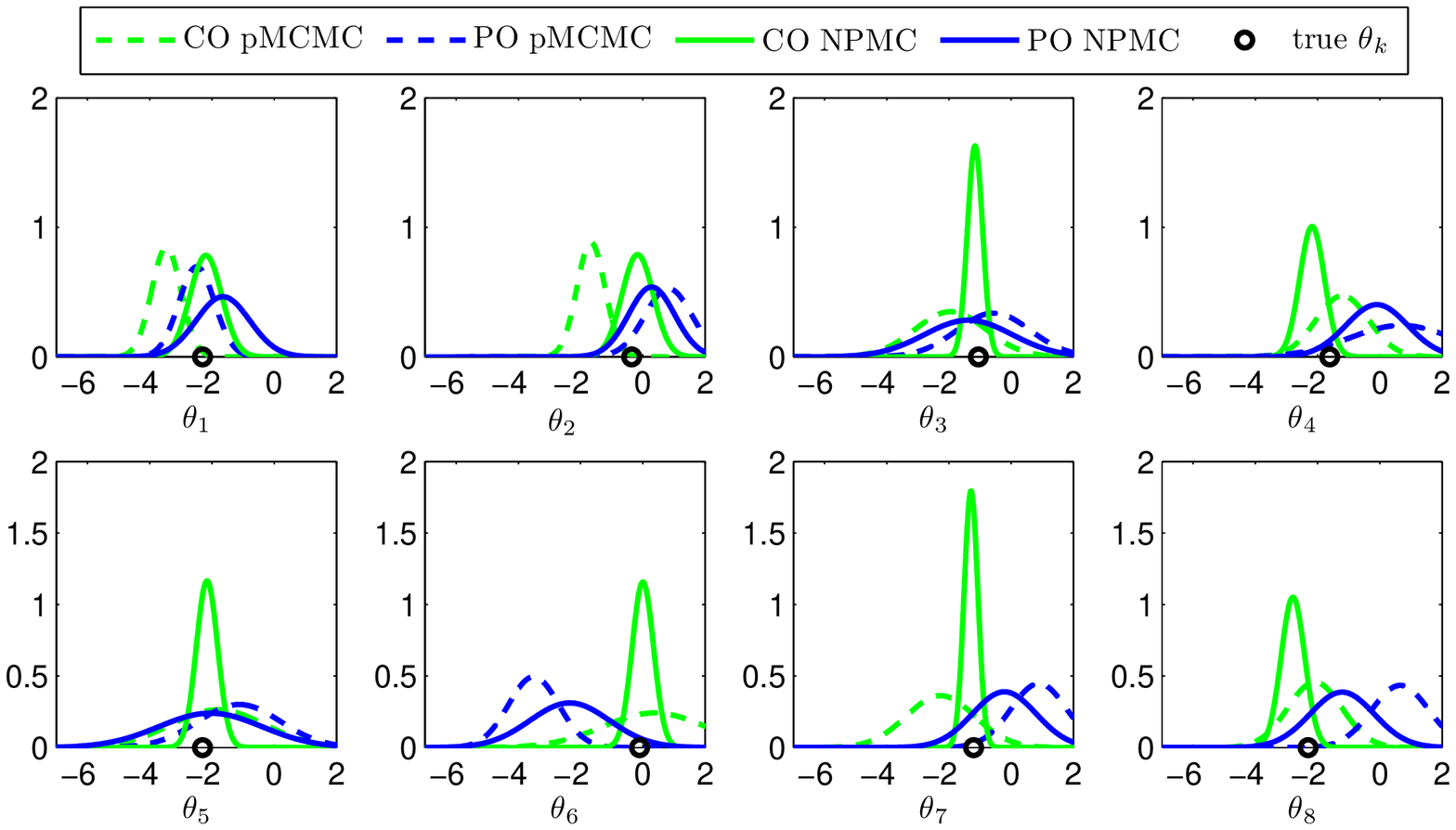}
\caption{Marginal posterior pdf approximations of each parameter
$\hat{p}(\theta_k|\textbf{y})$, $k=1,\ldots,K$, attained in an
average simulation run by the pMCMC and the NPMC, in the CO and in
the PO case.} \label{fig_theta_est}

\vspace{0.5cm}

\centering

\caption{Final MSE for the parameters $\theta_k$, $k = 1, \ldots, K$
in the CO and PO experiments, averaged over the simulation runs. The
last two columns corresponds to the mean and standard deviation
(std) values of the global MSE (averaged over the parameters). The
prior values are included for comparison} \label{tabla_8params}
\begin{tabular}{llllllllllll}
\hline\noalign{\smallskip}
  & & $\theta_1$ & $\theta_2$ & $\theta_3$ & $\theta_4$ & $\theta_5$ & $\theta_6$ & $\theta_7$ & $\theta_8$ & mean MSE & std MSE \\
\noalign{\smallskip}\hline\noalign{\smallskip}
  Prior & & 6.789 & 11.344 & 8.853 & 7.543 & 6.789 & 12.484 & 8.430 & 6.789 & 8.628 & 0\\
\noalign{\smallskip}\hline\noalign{\smallskip}

\multirow{2}{*}{PO} & pMCMC & 3.412 & 3.319 & 5.543 & 3.200 & 7.059
& 8.929 & 6.799 & 4.371 &
5.329 & 2.926 \\
& NPMC & 1.246 & 1.011 & 2.214 & 1.490 & 4.073 & 7.015 & 2.311 & 1.856 & 2.652 & 1.020\\
\multirow{2}{*}{CO} & pMCMC & 2.899 & 2.958 & 1.676 & 1.572 & 1.604
& 1.547 & 1.573 & 1.468 & 1.912 & 1.476 \\
& NPMC & 0.305 & 0.302 & 0.162 & 0.167 & 0.280 & 0.280 & 0.156 & 0.168 & 0.228 & 0.091 \\
\noalign{\smallskip}\hline\noalign{\smallskip}
\end{tabular}

\vspace{0.8cm}

\includegraphics[width=0.46\textwidth]{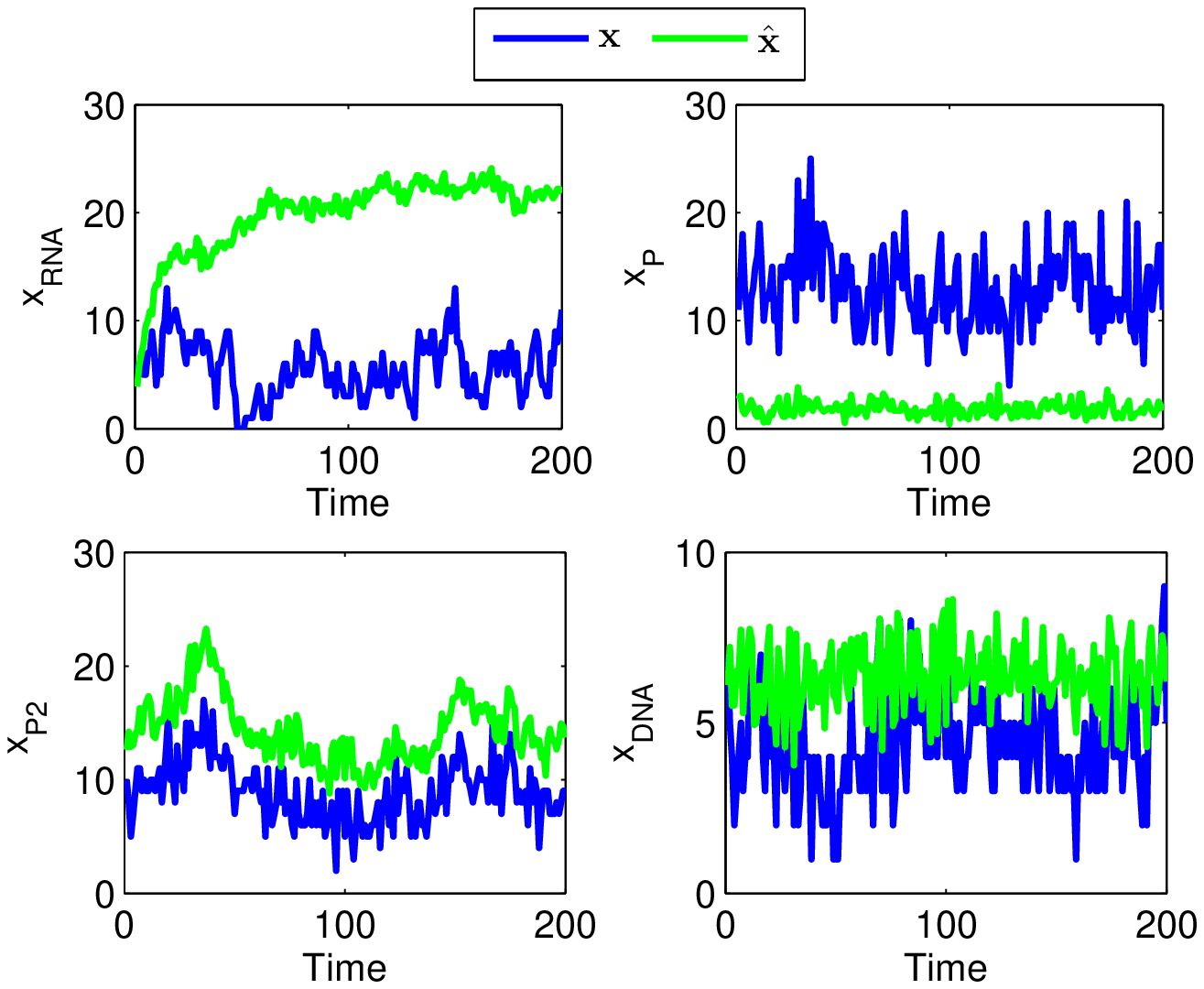}
\includegraphics[width=0.46\textwidth]{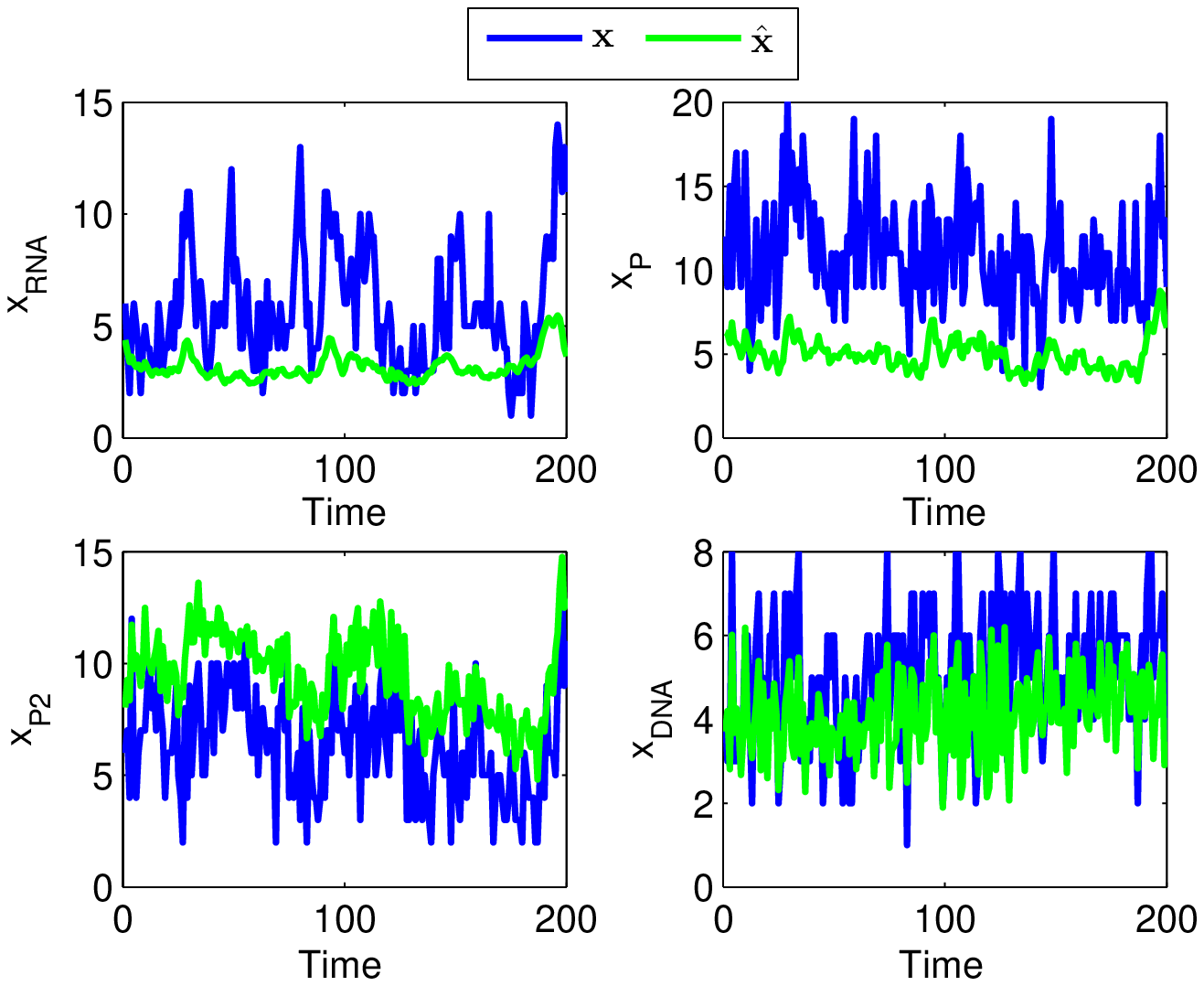}
\caption{Posterior mean $\hat{\textbf{x}} =
E_{p(\textbf{x}|\textbf{y})}[\textbf{x}]$ of the populations of all
species obtained in the average simulation run of the pMCMC
(\textit{left}) and the NPMC (\textit{right}) schemes, in the PO
scenario.} \label{fig_x_est}

\end{figure*}

Figure \ref{fig_8params_NPMC} depicts the evolution along the
iterations of the NESS (\textit{left}) and the MSE (\textit{right})
averaged over $P=100$ independent simulation runs for the NPMC
algorithm. Both indices converge to a steady value in a low number
of iterations also in this complex scenario. As expected, a
significantly higher final MSE is attained in the extremely data
poor PO scenario.

In Figure \ref{fig3_8params_pMCMC} (\textit{left}) the average ACF
attained by the pMCMC in the CO and the PO cases is represented.
Even after thinning the output, the sample correlation is extremely
high in both scenarios, which leads to a very low NESS. The
acceptance rate is also very low and very long chains are required
to obtain reasonable results. In the PO scenario 43.69 samples are
accepted on average in a simulation run of $I=15 \times 10^3$
samples (acceptance rate 0.0029). In the CO case, only 23.07 samples
are accepted on average (rate 0.0015).

Figure \ref{fig3_8params_pMCMC} (\textit{right}) depicts the final
Markov chain provided by the pMCMC method (after removing the
burn-in period and thinning the output) in the average simulation
run represented with a big square in Figure
\ref{fig_scatter_8params} (\textit{left}). It can be observed that
the mixing of the chain is very poor, with a total number of
accepted samples of 46 (close to the average). Many other
simulations, both in the PO and the CO scenarios, provide even lower
number of accepted samples, and thus, very inconsistent results.

Figure \ref{fig_theta_est} depicts the final Gaussian approximations
of the marginal posteriors $p(\theta_k | \textbf{y})$,
$k=1,\ldots,8$, obtained by the pMCMC and the NPMC methods, in the
CO and PO scenarios, for the average simulation runs represented as
big circles and squares in Figure \ref{fig_scatter_8params}. We can
observe that the NPMC method provides a significantly better
approximation of the log-rate parameters in the CO scenario, where a
larger amount of data is available, which is also clear from Figure
\ref{fig_scatter_8params} (\textit{right}). However, the pMCMC on
average performs similarly in both scenarios, due to the low
efficiency of the pMCMC sampling scheme when the dimension of the
problem (either $K$ or $N$) increases.

In Table \ref{tabla_8params} the MSE of each parameter $\theta_k$
averaged over $P=100$ independent simulation runs is shown, as
obtained with the pMCMC and the NPMC schemes, for the CO and the PO
experiments. In the CO case, NPMC provides homogeneous results for
all parameters. On the contrary, in the PO case, some of the
parameters (specially $\theta_5$ and $\theta_6$) are significantly
poorly estimated, presenting a final MSE close to the initial value
(which corresponds to the prior knowledge). The pMCMC scheme
presents significantly higher MSE values than NPMC in both
observation scenarios and for all parameters $\theta_k$.

Figure \ref{fig_x_est} depicts the population posterior mean
$\hat{\textbf{x}} = E_{p(\textbf{x}|\textbf{y})}[\textbf{x}]$
corresponding to the average simulation runs of the pMCMC and the
NPMC methods in the PO scenario, represented as big squares in
Figure \ref{fig_scatter_8params}. Again, the NPMC method provides
more accurate estimates of the unobserved populations than the pMCMC
method, specially for $x_{RNA}$. In the CO scenario both methods
provide good approximations of the populations of all species.

\section{Asymptotic convergence of NIS with approximate weights}
\label{Analysis}

\subsection{Scope of the analysis}

An analysis of the asymptotic effect of the transformation of the
weights on the IS-based approximation of integrals w.r.t. a target
probability distribution has already been addressed in
\citep{Koblents2013a}. In particular, the results in
\citep{Koblents2013a} show that, as long as $\frac{M_T}{M}\rw 0$,
the distortion introduced by the \textit{clipping} of the weights
vanishes asymptotically and the approximation of integrals of
bounded functions using IWs and using TIWs both converge to the same
value almost surely (a.s.). However,
\begin{itemize}
\item the argument in \citep{Koblents2013a} is based on classical concentration-of-measure inequalities and, therefore, rates are only found for convergence in probability, and
\item more importantly, the analysis relies on the ability to compute the non-normalized IWs exactly.
\end{itemize}
It is apparent from the algorithm description in Section
\ref{PMC_section} that, in the case of the SKM models of interest in
this paper, the IWs can only be approximated (via particle
filtering) and, therefore, the assumptions on which the theoretical
results of \citep{Koblents2013a} rely are not satisfied. In this
section, we improve on the analysis in \citep{Koblents2013a} by
looking explicitly into the convergence of the approximations of
integrals computed using approximate weights (both IWs and TIWs). We
provide convergence rates for the $L_p$ norms of the approximation
errors and show that the approximate weights computed by a standard
particle filter are ``good enough'' to ensure that these results
hold.

\subsection{Notation and basic assumptions}

Let $\pi(\bftheta)$ be the pdf associated to the target probability
distribution, let $q(\bftheta)$ be the importance function used to
propose samples in an IS scheme (not necessarily normalized) and let
$h(\bftheta) \propto \pi(\bftheta)$ be a function proportional to
$\pi$, with the proportionality constant independent of $\bftheta$.
The samples drawn from the distribution associated to $q$ are
denoted $\bftheta^{(i)}$, $i=1, ..., M$, and their associated
non-normalized IWs are $w^{(i)*} = h( \bftheta^{(i)} ) / q (
\bftheta^{(i)})$, $i=1, ..., M$.

Let us define the weight function $g(\bftheta) =
h(\bftheta)/q(\bftheta)$ and, in particular, $g( {\bftheta}^{(i)}) =
w^{(i)*}$. The support of $g$ is the same as the support of $q$,
denoted $\sS \subseteq \mathbb{R}^K$. If we assume that both
$q(\bftheta)>0$ and $\pi(\bftheta) > 0$ for any $\bftheta \in \sS$,
then $g(\bftheta) > 0$ for every $\bftheta \in \sS$ as well. Also,
trivially, $\pi \propto gq$, with the proportionality constant
independent of $\bftheta$. These assumptions are standard for
classical IS.

Assume that the standard IWs can be computed exactly. In that case,
the approximation $\pi^M$ of the target probability measure can be
written as
\begin{equation*}
\pi^M(d\bftheta) = \sum_{i=1}^M w^{(i)}
\delta_{{\bftheta}^{(i)}}(d\bftheta),
\end{equation*}
where $w^{(i)} = \frac{
    g ( {\bftheta}^{(i)} )
}{
    \sum_{j=1}^M g ( {\bftheta}^{(j)} )
}$, $i=1, ..., M$.

Assume next that the weight function cannot be evaluated exactly
but, instead, a sequence of approximations $g^J(\bftheta)$, $J \in
\mbN$, exists for any point $\bftheta \in \sS$. We denote the random
measure constructed from the approximate IWs as
\begin{equation*}
\pi^{M,J}(d\bftheta) = \sum_{i=1}^M w^{(i),J}
\delta_{{\bftheta}^{(i)}}(d\bftheta),
\end{equation*}
where $w^{(i),J} = \frac{
    g^J ( {\bftheta}^{(i)} )
}{
    \sum_{j=1}^M g^J ( {\bftheta}^{(j)} )
}$, $i=1, ..., M$. Let us denote by $\varphi^M$ the nonlinear
transformation function used to compute non-normalized TIWs, i.e.,
$\bar{w}^{(i)*} = \varphi^M ( w^{(i)*} )$, $i=1,\ldots, M$, where
$w^{(i)*}$ is the standard unnormalized IW associated to the sample
$\bftheta^{(i)}$.
Then the weighted approximation of $\pi(\bftheta)d\bftheta$
constructed according to the NIS scheme is
\begin{equation*}
\bar \pi^{M,J}(d\bftheta) = \sum_{i=1}^M \bar w^{(i),J}
\delta_{{\bftheta}^{(i)}}(d\bftheta),
\end{equation*}
where $\bar w^{(i),J} = \frac{\varphi^M ( g^J ( {\bftheta}^{(i)}
))}{\sum_{j=1}^M \varphi^M ( g^J ( {\bftheta}^{(j)}))} $, $i=1, ...,
M$.

We make the following assumptions on the transformation function
$\varphi^M$, the weight function $g$ and its approximations $\{ g^J:
J \ge 1\}$.
\begin{itemize}
\item[A1]  The nonlinear transformation $\varphi^M$ of the weights is of a
\emph{clipping} class. In particular, given an index permutation
$i_1, \ldots, i_M$ such that $w^{(i_1)*} \geq \ldots \geq
w^{(i_M)*}$, and a choice of the \textit{clipping} parameter $M_T <
M$, the transformation $\varphi^M$ can be expressed as\footnote{Note
that $\varphi^M$ is a function of both the complete weight set
$\{w^{(j)*}\}_{j=1}^M$ and the index of the weight to be
transformed, i.e., $\varphi^M:\{ w^{(j)*}, j=1,\ldots,M\} \times \{
1,\ldots,M \} \rw [1,+\infty)$.}
\begin{equation*}
\label{eqConditionPhi} \varphi^M ( w^{(i_k)*} ) \! =
 \! \left\{
    \begin{array}{ll}
    w^{(i_{M_T})*}, & \mbox{for } k=1, \ldots, M_T, \; \mbox{and}\\
    w^{(i_k)*}, & \mbox{for } k=M_T+1, \ldots, M.
    \end{array}
\right..
\end{equation*}

\item[A2] The weight function $g$ has a finite upper bound and a positive lower bound. Specifically, there exists a real number $0 < a < \infty$ such that
$ a^{-1} \le g(\bftheta) \le a $ for every $\bftheta \in \sS$.

\item[A3] The same bounds of the weight function $g$ hold for its approximations $g^J$, $J \ge 1$. To be specific, the inequalities
$ a^{-1} \le g^J(\bftheta) \le a $ hold for every $\bftheta \in
\sS$, any $J \ge 1$ and the same real number $0<a<\infty$ as in A2.

\item[A4] The approximation $g^J$ of the weight function is possibly random and satisfies the inequality
$$
\sup_{\bftheta \in \sS} | g(\bftheta) - g^J(\bftheta) | \le
\frac{W_{g,\epsilon}}{J^{\frac{1}{2}-\epsilon}}
$$
where $W_{g,\epsilon}$ is a positive a.s. finite random variable and
$0 < \epsilon < \frac{1}{2}$ is an arbitrarily small constant, both
independent of $J$.

\end{itemize}

Note that if the support set $\sS$ is compact then assumption A2
holds whenever $q>0$ and $h>0$ in $\sS$. Otherwise, the proposal $q$
has to be chosen so that it has heavier tails than $\pi$.

In the sequel we look into the approximation of integrals of the
form $(f,\pi) = \int I_{\sS}(\bftheta) f(\bftheta) \pi(\bftheta)
d\bftheta$, where $I_\sS(\bftheta)$ is an indicator
function\footnote{Namely, $I_\sS(\bftheta)=1$ if $\bftheta \in \sS$
and $I_\sS(\bftheta)=0$ otherwise.} and $f$ is a bounded real
function in the parameter space $\sS$. We use $\|f\|_\infty =
\sup_{\bftheta \in \sS}|f(\bftheta)| < \infty$ to denote the
supremum norm of a bounded function. The set of bounded functions on
$\sS$ is $B(\sS) = \{ f: \sS \rw \mathbb{R} : \|f\|_\infty <
\infty\}$. The approximations of interest are
\begin{eqnarray}
(f,\pi^{M,J}) &=& \sum_{i=1}^M f(\bftheta^{(i)}) w^{(i),J}, \quad \mbox{and} \nonumber\\
(f,\bar\pi^{M,J}) &=& \sum_{i=1}^M f(\bftheta^{(i)}) \bar w^{(i),J}.
\nonumber
\end{eqnarray}

\subsection{Convergence rates}

The following basic Lemma establishes that both $(f,\bar \pi^{M,J})$
and $(f,\bar \pi^{M,J})$ converge toward $(f,\pi)$ a.s. and provides
explicit rates for the absolute approximation errors.

\begin{lemma} \label{lmBasic}
Assume that A1, A2, A3 and A4 hold,
$$
J=J(M) \ge M \quad \mbox{and} \quad M_T \le \sqrt{M}.
$$
Then, there exist positive and a.s. finite random variables
$W_{f,g,\epsilon}$ and $\bar W_{f,g,\epsilon}$, independent of $M$
and $J$, such that
\begin{equation}
| (f,\pi^{M,J}) - (f,\pi) | \le \frac{
    W_{f,g,\epsilon}
}{
    M^{\frac{1}{2}-\epsilon}
} \label{eq1Basic}
\end{equation}
and
\begin{equation}
| (f,\bar \pi^{M,J}) - (f,\pi) | \le \frac{
    \bar W_{f,g,\epsilon}
}{
    M^{\frac{1}{2}-\epsilon}
} \label{eq2Basic}
\end{equation}
for every $f \in B(\sS)$, where $0<\epsilon<\frac{1}{2}$ is an
arbitrarily small constant independent of $M$ and $J$. In particular
\begin{equation}
\lim_{M\rw\infty}  (f,\pi^{M,J}) = \lim_{M\rw\infty} (f,\bar
\pi^{M,J}) = (f,\pi) \mbox{ a.s.} \label{eqLemmaLimits}
\end{equation}
\end{lemma}

A proof is provided in Appendix \ref{apLemmaBasic}. Lemma
\ref{lmBasic} shows that we attain the usual Monte Carlo rate of
convergence ($M^{-\frac{1}{2}+\epsilon}$) despite the approximation
of the IWs and its subsequent \textit{clipping} to compute TIWs.
Note, however, that the random variables $W_{f,g,\epsilon}$ and
$\bar W_{f,g,\epsilon}$ are not equal and, in general,
$W_{f,g,\epsilon} \le \bar W_{f,g,\epsilon}$.

\subsection{Approximate weights via particle filtering}

In this section we introduce a more precise notation for the
state-space model (compared to the argument-wise used in the
previous sections), in order to perform the analysis with
approximate weights. Assume we have a discrete-time state space
Markov model with state process $\{ \textbf{X}_n \}_{n\ge 0}$ taking
values on $\mX \subseteq \Real^{d_\textbf{x}}$ and an observation
process $\{ \textbf{Y}_n \}_{n \ge 0}$ taking values on $\mY
\subseteq \Real^{d_\textbf{y}}$. The prior distribution (probability
measure) of the state is now denoted $\tau_0(d\textbf{x})$ and the
transition (Markov) kernel depends on a vector-valued random
parameter $\boldsymbol{\Theta}$ that takes values on a compact set
$\sS \subset \Real^{d_{\bftheta}}$ and has prior distribution
$\mu_0(d\bftheta)$ independent of $\textbf{X}_0$. In particular, the
Markov kernel is now denoted
$\tau_{n,\bftheta}(d\textbf{x}_n|\textbf{x}_{n-1})$ and the
conditional density of the observations is
$u_n(\textbf{y}_n|\textbf{x}_n)>0$. The latter also yields the
likelihood of the signal $\textbf{x}_n$, hence we often write, for
conciseness, $u_n^{\textbf{y}_n}(\textbf{x}_n) \dfn
u_n(\textbf{y}_n|\textbf{x}_n)$.

At time $n$, the one-step-ahead predictive distribution of the state
$\textbf{X}_n$ given fixed observations
$\textbf{Y}_{1:n-1}=\textbf{y}_{1:n-1}$ and a parameter value
$\boldsymbol{\Theta}=\bftheta$ is denoted $\xi_{n,\bftheta}$,
specifically, for any Borel subset $A \subset \mX$,
$$
\xi_{n,\bftheta}(A) = \mbP_n\left( \textbf{X}_n \in A |
\textbf{Y}_{1:n-1}=\textbf{y}_{1:n-1}, \boldsymbol{\Theta} =
\bftheta \right) \footnote{$\mbP_n$ denotes the joint probability
measure for the set of random variables $\{ \textbf{x}_k \}_{k\leq
n} \cup \{\textbf{y}_k \}_{k\leq n} \cup \{ \boldsymbol{\Theta} \}$
on the measurable space $(\sigma(\textbf{x}_{0:n}, \textbf{y}_{1:n},
\boldsymbol{\Theta}), \mX^{n+1} \times \mathcal{Y}^n \times \sS)$.}.
$$
The filter measure at time $n$ given observations
$\textbf{Y}_{1:n}=\textbf{y}_{1:n}$ and parameter
$\boldsymbol{\Theta}=\bftheta$ is denoted $\phi_{n,\bftheta}$,
namely,
$$
\phi_{n,\bftheta}(A) = \mbP_n\left( \textbf{X}_n \in A |
\textbf{Y}_{1:n}=\textbf{y}_{1:n}, \boldsymbol{\Theta} = \bftheta
\right).
$$
The predictive measure $\xi_{n,\bftheta}$ can be expressed in terms
of $\tau_{n,\bftheta}$ and $\phi_{n-1,\bftheta}$. Specifically, we
write $\xi_{n,\bftheta}=\tau_{n,\bftheta}\phi_{n-1,\bftheta}$,
meaning that, for any integrable function $f : \mX \rw \Real$,
\begin{eqnarray}
(f,\xi_{n,\bftheta}) &=& \int \int f(\textbf{x})\tau_{n,\bftheta}(d\textbf{x}|\textbf{x}') \phi_{n-1,\bftheta}(d\textbf{x}') \nonumber\\
&=& (f,\tau_{n,\bftheta}\phi_{n-1,\bftheta}). \nonumber
\end{eqnarray}
We also note that
$$
(f,\xi_{n,\bftheta}) = (\bar f_n,\phi_{n-1,\bftheta}),
$$
where $\bar f_n(\textbf{x}') = \int
f(\textbf{x})\tau_{n,\bftheta}(d\textbf{x}|\textbf{x}')$. The filter
measures $\phi_{n,\bftheta}$ and $\phi_{n-1,\bftheta}$ are related
by the projective product
$$
\phi_{n,\bftheta} = u_n^{\textbf{y}_n} \star
\tau_{n,\bftheta}\phi_{n-1,\bftheta} = u_n^{\textbf{y}_n} \star
\xi_{n,\bftheta},
$$
defined as \citep{Bain2008}
$$
(f,u_n^{\textbf{y}_n} \star \xi_{n,\bftheta}) \dfn \frac{
    (fu_n^{\textbf{y}_n}, \xi_{n,\bftheta})
}{
    (u_n^{\textbf{y}_n},\xi_{n,\bftheta})
}.
$$


Let
$$
\xi_{n,\bftheta}^J(d\textbf{x}) = \frac{1}{J} \sum_{j=1}^J
\delta_{\textbf{x}_n^{(j)}}(d\textbf{x}) \mbox{ and }
$$
$$
\phi_{n,\bftheta}^J(d\textbf{x}) = \frac{1}{J} \sum_{j=1}^J
\delta_{\tilde{\textbf{x}}_n^{(j)}}(d\textbf{x})
$$
be the approximations of $\xi_{n,\bftheta}$ and $\phi_{n,\bftheta}$
produced by a standard particle filter \citep{Gordon1993} with $J$
particles. We have the following theoretical guarantee for the
convergence of $\xi_{n,\bftheta}^J$ and $\phi_{n,\bftheta}^J$.
\begin{lemma} \label{lmUniformPF}
Let $N$ be a finite time horizon and let
$\textbf{Y}_{1:N}=\textbf{y}_{1:N}$ be an arbitrary but fixed
sequence of observations. Assume that, for every $n=1, ..., N$,
$u_n^{\textbf{y}_n} \in B(\mX)$, $\sS$ is compact and
\begin{equation}
\inf_{\bftheta \in \sS} (u_n^{\textbf{y}_n},\xi_{n,\bftheta}) > 0.
\label{asInf}
\end{equation}
Then, for every $f \in B(\mX)$, every $p \ge 1$ and every $n=0, 1,
..., N$,
\begin{eqnarray}
\sup_{\bftheta \in \sS} \| (f,\xi_{n,\bftheta}^J) -
(f,\xi_{n,\bftheta}) \|_p &\le& \frac{
    c_{1,n} \| f \|_\infty
}{
    \sqrt{J}
} \label{eqUCPF-1} \\
\sup_{\bftheta \in \sS} \| (f,\phi_{n,\bftheta}^J) -
(f,\phi_{n,\bftheta}) \|_p &\le& \frac{
    c_{2,n} \| f \|_\infty
}{
    \sqrt{J}
}, \label{eqUCPF-2}
\end{eqnarray}
where $c_{1,n}$ and $c_{2,n}$ are positive and finite constants
independent of $J$ and $\bftheta$.
\end{lemma}
\indent {\bf Proof.} This is a straightforward consequence of
\cite[Lemma 2]{Crisan2013nested}. 
$\qed$

We denote the likelihood of the parameter realization $\bftheta$
given the observations $\textbf{Y}_{1:N} = \textbf{y}_{1:N}$ as
$\lambda_N(\bftheta)$, where
$$
\lambda_N(\bftheta) \dfn \prod_{n=1}^N
(u_n^{\textbf{y}_n},\xi_{n,\bftheta})
$$
(it is straightforward to show that $\lambda_N(\bftheta)$ yields the
value of the joint pdf of $\textbf{y}_1, \ldots, \textbf{y}_N$
conditional on $\bftheta$). This likelihood can be naturally
approximated via particle filtering as
$$
\lambda_N^J(\bftheta) \dfn \prod_{n=1}^N
(u_n^{\textbf{y}_n},\xi_{n,\bftheta}^J)
$$
and still guarantee that $\lambda_N^J \rw \lambda_N$ a.s. with
standard Monte Carlo rates. This is rigorously stated below.
\begin{lemma} \label{lmUniformL}
Under the assumptions of Lemma \ref{lmUniformPF} there exists a
positive and a.s. finite random variable $W_{N,u,\epsilon}$
independent of $J$ such that
\begin{equation}
\sup_{\bftheta \in \sS} | \lambda_N^J(\bftheta) -
\lambda_N(\bftheta) | \le \frac{
    W_{N,u,\epsilon}
}{
    J^{\frac{1}{2}-\epsilon}
}, \label{eqConvL_J}
\end{equation}
where $0<\epsilon< \frac{1}{2}$ is an arbitrarily small constant
independent of $J$. In particular, the inequality \eqref{eqConvL_J}
implies that $\lim_{J\rw\infty} \lambda_N^J(\bftheta) =
\lambda_N(\bftheta)$ a.s. and uniformly over $\bftheta \in \sS$.
\end{lemma}
\noindent {\bf Proof.} See Appendix \ref{apProofUniformL}. $\qed$

\subsection{Convergence of the NIS scheme with approximate weights}

We can put the previous Lemmas together to prove convergence of the
NIS scheme with approximate weights.

Assume that we use NIS to approximate the posterior measure of the
parameter $\boldsymbol{\bftheta}$, namely
\begin{equation}
\pi(\bftheta)d\bftheta = \mbP_N\left( \boldsymbol{\Theta} \in
d\bftheta | \textbf{Y}_{1:N} = \textbf{y}_{1:N} \right).
\label{eqTargetPost}
\end{equation}
It is straightforward to show that
\begin{equation}
\pi(\bftheta) \propto h(\bftheta) = \lambda_N(\bftheta)
m_0(\bftheta), \nonumber
\end{equation}
where $m_0(\bftheta)$ is the density associated to the prior
probability distribution of the parameter, $\mu_0$. If a proposal
pdf $q$ is used, the weight function becomes
\begin{equation}
g(\bftheta) = \frac{
    h(\bftheta)
}{
    q(\bftheta)
} = \frac{
    \lambda_N(\bftheta) m_0(\bftheta)
}{
    q(\bftheta)
}. \nonumber
\end{equation}
Since the likelihood $\lambda_N(\bftheta)$ cannot be computed in
closed form we readily approximate it using a particle filter. This,
in turn, yields the approximate weight function
\begin{equation}
g^J(\bftheta) = \frac{
    h^J(\bftheta)
}{
    q(\bftheta)
} = \frac{
    \lambda_N^J(\bftheta) m_0(\bftheta)
}{
    q(\bftheta)
}. \label{eqWeightFunction_via_PF}
\end{equation}

Let us apply a NIS scheme to approximate the target distribution in
\eqref{eqTargetPost}, where the weight function can be approximately
evaluated using \eqref{eqWeightFunction_via_PF}. The approximation
of $\pi$ with standard IWs is denoted $\pi^{M,J}$ and the
approximation with TIWs is denoted $\bar \pi^{M,J}$. The
observations $\textbf{y}_{1:N}$ are arbitrary but fixed. Then we
have the following result.

\begin{theorem} \label{thNIS_via_PF}
Assume that A1 holds, $J=J(M)\ge M$, $M_T\le M$, $u_n^{\textbf{y}_n}
\in B(\mX)$ for every $n=1,\ldots,N$ and there exists a real
constant $a>0$ such that $\inf_{\textbf{x}\in \mX}
u_n^{\textbf{y}_n} \ge \frac{1}{a}$ for every $n=1, ..., N$. If the
inequalities
\begin{eqnarray}
\| m_0/q \|_\infty = \sup_{\bftheta \in \sS} \frac{
    m_0(\bftheta)
}{
    q(\bftheta)
} &<& \infty,
\label{eqAssumptionSup} \\
\mbox{and } \inf_{\bftheta \in \sS} \frac{
    m_0(\bftheta)
}{
    q(\bftheta)
} &>& 0 \nonumber
\end{eqnarray}
are satisfied, then, for every $f \in B(\sS)$, there exist positive
random variables $W_{f,g,\epsilon}$ and $\bar W_{f,g,\epsilon}$,
a.s. finite and independent of $M$ and $J$, such that
\begin{eqnarray}
| (f,\pi^{M,J}) - (f,\pi) | &\le& \frac{
    W_{f,g,\epsilon}
}{
    M^{\frac{1}{2}-\epsilon}
}, \quad \mbox{and} \label{eqTeorIneq1} \\
| (f,\bar \pi^{M,J}) - (f,\pi) | &\le& \frac{
    \bar W_{f,g,\epsilon}
}{
    M^{\frac{1}{2}-\epsilon}
}, \label{eqTeorIneq2}
\end{eqnarray}
where $0<\epsilon<\frac{1}{2}$ is an arbitrarily small constant
independent of $M$. The inequalities \eqref{eqTeorIneq1} and
\eqref{eqTeorIneq2} imply
\begin{equation}
\lim_{M\rw\infty} (f,\pi^{M,J}) = \lim_{M\rw\infty}  (f,\bar
\pi^{M,J}) = (f,\pi)  \quad \textrm{a.s.} \nonumber
\end{equation}
\end{theorem}
\noindent{\bf Proof.} The absolute error in the approximation of the
weight function is
\begin{equation}
| g(\bftheta) - g^J(\bftheta) | = \frac{m_0(\bftheta)}{q(\bftheta)}
| \lambda_N^J(\bftheta) - \lambda_N(\bftheta) |. \label{eqIneqObvia}
\end{equation}
However, from Lemma \ref{lmUniformL}, we readily have\footnote{The
assumptions of Theorem 1 imply the assumptions of Lemmas 2 and 3. In
particular, $\inf_{\textbf{x} \in \mX} u_n^{\textbf{y}_n} \geq
\frac{1}{a}$ implies $\inf_{\bftheta \in \sS} (u_n^{\textbf{y}_n},
\xi_{n,\bftheta}) > 0$.}
\begin{equation}
\sup_{\bftheta \in \sS} | \lambda_N^J(\bftheta) -
\lambda_N(\bftheta) | \le \frac{
    W_{N,u,\epsilon}
}{
    J^{\frac{1}{2}-\epsilon}
} \label{eqTarara}
\end{equation}
where $W_{N,u,\epsilon}>0$ is a.s. finite and $0 < \epsilon <
\frac{1}{2}$ is arbitrarily small, and both are independent of $J$
(and $M$). Substituting \eqref{eqTarara} and \eqref{eqAssumptionSup}
into \eqref{eqIneqObvia} yields
\begin{equation}
\sup_{\bftheta \in \sS} | g(\bftheta) - g^J(\bftheta) | \le \frac{
    W_{N,u,\epsilon} \| m_0/q \|_\infty
}{
    J^{\frac{1}{2}-\epsilon}
} \nonumber
\end{equation}
and, as a consequence, the sequence of approximate weight functions
$g^J$ satisfies A4 with
$$
W_{g,\epsilon} = \| m_0/q \|_\infty W_{N,u,\epsilon} > 0
$$
a.s. finite.

Assumptions A2 and A3 are also satisfied. In particular, since
$u_n^{\textbf{y}_n}\in B(\mX)$ for every $n = 1, ..., N$, it follows
that
$$
\prod_{n=1}^N (u_n^{\textbf{y}_n},\alpha) \le \prod_{n=1}^N \|
u_n^{\textbf{y}_n} \|_\infty < \infty
$$
for any probability measure on $(\mB(\mX),\mX)$ (where $\mB(\mX)$
denotes the Borel $\sigma$-algebra of subsets of $\mX$). In
particular, $\prod_{n=1}^N \| u_n^{\textbf{y}_n} \|_\infty$ is an
upper bound for $\lambda_N$ and $\lambda_N^J$. Moreover, since
$\inf_{x\in \mX} u_n^{\textbf{y}_n} \ge a^{-1}$ for every $n=1, ...,
N$ it follows that
$$
\prod_{n=1}^N (u_n^{\textbf{y}_n},\alpha) \ge a^{-N} > 0
$$
for any probability measure $\alpha$ on $(\mB(\mX),\mX)$. In
particular, $a^{-N}$ is a positive lower bound for both $\lambda_N$
and $\lambda_N^J$. The factor $m_0/q$, independent of the
approximation index $J$, has a positive lower bound and a finite
upper bound by assumption.

Since A1--A4 are satisfied, we can apply Lemma \ref{lmBasic}, which
yields \eqref{eqTeorIneq1} and \eqref{eqTeorIneq2} directly. $\qed$


\section{Conclusion}
\label{Conclusion}

We have addressed the problem of approximating posterior
distributions of the parameters and the populations in stochastic
kinetic models. We have applied a nonlinear population Monte Carlo
(NPMC) method, which iteratively approximates the target
distribution via an importance sampling scheme. The NPMC method
resorts to a sequential Monte Carlo approximation of the posterior
populations to evaluate the importance weights. Additionally, it
performs nonlinear transformations to the weights to avoid
degeneracy and the numerical problems typically arising in the
proposal update of the PMC scheme in high dimensional problems. We
provide an extended convergence analysis of the nonlinear importance
sampling scheme, which takes into account the weight approximation.

We have compared the performance of the NPMC method to the well
known particle Markov chain Monte Carlo (pMCMC) method, applied to
the challenging prokaryotic autoregulatory model. Both methods have
been applied in the exact simulation form, since the complexity of
this model allows to do so. We show how the NPMC method outperforms
the pMCMC method and requires only a moderate computational cost.
Besides, the proposed method has a set of important features, common
to all PMC schemes, as the sample independence, ease of
parallelization, and compared to MCMC schemes, there is no need for
convergence (burn-in) periods.

\appendix

\section{Sequential Monte Carlo approximation of $p(\textbf{x} | \bftheta, \textbf{y})$ and $p(\textbf{y} | \bftheta)$}
\label{apPF}

In this appendix we provide details on the approximation of the
posterior $p(\textbf{x} | \bftheta, \textbf{y})$ and the likelihood
$p(\textbf{y}|\bftheta)$. For a given vector of log-rate parameters
$\bftheta$, the following standard particle filter (see, e.g.,
\citep{Doucet2001}) is run.

\vspace{0.3cm}

\underline{\textbf{Initialization ($n=0$}):}

Draw a collection of $J$ samples $\{ \textbf{x}_0^{(j)} \}_{j=1}^{J}
\sim p(\textbf{x}_0)$.

\vspace{0.2cm}

\underline{\textbf{Recursive step ($n=1, \ldots, N$)}:}
\begin{enumerate}
\item Draw  $\{ \textbf{x}_{n}^{(j)} \}_{j=1}^{J} \sim p ( \textbf{x}_{n} | \textbf{x}_{n-1}^{(j)}, \bftheta )$ using
the Gillespie algorithm (or a diffusion approximation).

\item Construct $\textbf{x}_{1:n}^{(j)} = [{\textbf{x}_{1:n-1}^{(j)}}^\top,
{\textbf{x}_{n}^{(j)}}^\top]^\top$.

\item Compute normalized IWs
$\omega_{n}^{(j)*} = p ( \textbf{y}_{n} | \textbf{x}_{n}^{(j)} )$,
$\omega_{n}^{(j)} = \omega_{n}^{(j)*}/\sum_{l=1}^J
\omega_{n}^{(l)*}$, $j=1, \ldots, J$.

\item Resample $J$ times with replacement from $\{ \textbf{x}_{1:n}^{(j)}
\}_{j=1}^{J}$ according to the weights $\{ \omega_{n}^{(j)}
\}_{j=1}^{J}$ to yield $\{\tilde{\textbf{x}}_{1:n}^{(j)}\}_{j=1}^J$.
\end{enumerate}

\vspace{0.3cm}

An approximation of the posterior $p(\textbf{x} | \bftheta,
\textbf{y})d\textbf{x}$ may be constructed from the final set of
samples $\textbf{x}_{1:N}^{(j)} = \textbf{x}^{(j)}$ and weights
$\omega_{N}^{(j)}$ as the discrete random measure
\begin{equation*}
\hat{p}^J(d\textbf{x} | \bftheta, \textbf{y}) = \sum_{j=1}^J
\omega_{N}^{(j)} \delta_{\textbf{x}^{(j)}} (d\textbf{x}).
\end{equation*}

The likelihood $p(\textbf{y}|\bftheta)$ can be approximated in turn
as
\begin{equation*}
\hat p^J(\textbf{y}|\bftheta) = \prod_{n=1}^N \frac{1}{J}
\sum_{j=1}^J p(\textbf{y}_n|\textbf{x}_n^{(j)}).
\end{equation*}

In order to obtain a sample from the approximation
$\hat{p}^J(d\textbf{x} | \bftheta, \textbf{y})$ in the pMCMC or the
NPMC schemes, we just draw a sample out of the set $\{
\textbf{x}^{(j)}\}_{j=1}^J$ according to their IWs $\omega_N^{(j)}$.


\section{Proof of Lemma \ref{lmBasic}} \label{apLemmaBasic}

We look into $(f,\pi^{M,J})$ first. Since
\begin{equation}
(f,\pi) = \frac{
    (fg,q)
}{
    (g,q)
} \mbox{ and } (f,\pi^{M,J}) = \frac{
    (fg^J,q^M)
}{
    (g^J,q^M)
}, \label{eqNormalizada0}
\end{equation}
where $q^M = \frac{1}{M} \sum_{i=1}^M \delta_{\bftheta^{(i)}}$, it
is simple to show that
\begin{eqnarray}
(f,\pi^{M,J}) - (f,\pi) &=& \frac{
    (fg^J,q^M) - (fg,q)
}{
    (g,q)
} \nonumber \\
&& + (f,\pi) \frac{
    (g,q) - (g^J,q^M)
}{
    (g,q)
}. \label{eqDecom}
\end{eqnarray}
However, since $(g,q)=(1,h) = \int
I_\sS(\bftheta)h(\bftheta)d\bftheta$ and $(f,\pi) \le \| f
\|_\infty$, Eq. \eqref{eqDecom} readily yields
\begin{eqnarray}
| (f,\pi^{M,J}) - (f,\pi) | &\le& \frac{
    1
}{
    (1,h)
} \left|
    (fg^J,q^M) - (fg,q)
\right| \nonumber \\
&& + \frac{
    \| f \|_\infty
}{
    (1,h)
} \left|
    (g,q) - (g^J,q^M)
\right|, \label{eqInicial}
\end{eqnarray}
and, therefore, the problem reduces to computing bounds for errors
of the form $| (bg^J,q^M) - (bg,q) |$, where $b \in B(\sS)$.

Choose any $b \in B(\sS)$. A simple triangle inequality yields
\begin{equation}
| (bg^J,q^M) - (bg,q) | \le | (bg^J,q^M) - (bg,q^M) | + | (bg,q^M) -
(bg,q) |. \label{eqTriang1}
\end{equation}
Since $q^M = \frac{1}{M} \sum_{i=1}^M \delta_{\bftheta^{(i)}}$, for
the second term on the right hand side of \eqref{eqTriang1} we can
write
\begin{equation}
\mbE\left[
    | (bg,q^M) - (bg,q) |^p
\right] = \mbE\left[
    \left|
        \frac{1}{M} \sum_{i=1}^M Z^{(i)}
    \right|^p
\right], \label{eqTriang2t}
\end{equation}
where the random variables
\begin{equation}
Z^{(i)} = b(\bftheta^{(i)})g(\bftheta^{(i)}) - (bg,q), \quad i=1,
..., M, \nonumber
\end{equation}
are i.i.d. with zero mean (since the $\bftheta^{(i)}$'s are i.i.d.
draws from $q$). Therefore, it is straightforward to show that
\begin{equation}
 \mbE\left[
    \left|
        \frac{1}{M} \sum_{i=1}^M Z^{(i)}
    \right|^p
\right] \le \frac{
    \tilde c^p a^p \| b \|_\infty^p
}{
    M^\frac{p}{2}
}, \label{eqZygmund}
\end{equation}
where $\tilde c$ is a constant independent of $M$ and $q$, and $a$
is the uniform upper bound for the weight function $g$ provided by
assumption A2, also independent of $M$. Combining \eqref{eqZygmund}
with \eqref{eqTriang2t} readily yields
\begin{equation}
\| (bg,q^M) - (bg,q) \|_p \le \frac{
    \tilde c a \| b \|_\infty
}{
    \sqrt{M}
}. \label{eq_tri2term}
\end{equation}
The inequality \eqref{eq_tri2term} implies that there exists an a.s.
finite random variable $U_\epsilon>0$ such that
\begin{equation}
| (bg,q^M) - (bg,q) | \le \frac{
    U_\epsilon
}{
    M^{\frac{1}{2}-\epsilon}
}, \label{eq_tri2term_eps}
\end{equation}
where $0 < \epsilon < \frac{1}{2}$ is an arbitrarily small constant
independent of $M$ (see \cite[Lemma 1]{Crisan2011particle}).

Expanding now the first term on the right hand side of
\eqref{eqTriang1} we find that
\begin{eqnarray}
\left|
    (bg^J,q^M) - (bg,q^M)
\right| &=& \left|
    \frac{1}{M} \sum_{i=1}^M b(\bftheta^{(i)}) \left(
        g^J(\bftheta^{(i)}) - g(\bftheta^{(i)})
    \right)
\right| \nonumber \\
&\le& \frac{\| b \|_\infty^p}{M} \sum_{i=1}^M \left|
    g^J(\bftheta^{(i)}) - g(\bftheta^{(i)})
\right|. \label{eqIneq2t}
\end{eqnarray}
However, by assumption A4, there exists an a.s. finite random
variable $W_{g,\epsilon}$ such that
\begin{equation}
\sup_{\bftheta \in \sS} \left|
    g^J(\bftheta) - g(\bftheta)
\right| \le \frac{
    W_{g,\epsilon}
}{
    J^{\frac{1}{2}-\epsilon}
}, \label{eqFromA4}
\end{equation}
where $0 <\epsilon < \frac{1}{2}$ is an arbitrary small constant
independent of $J$. Combining \eqref{eqFromA4} with \eqref{eqIneq2t}
yields
\begin{equation}
\left|
    (bg^J,q^M) - (bg,q^M)
\right| \le \frac{
    \| b \|_\infty W_{g,\epsilon}
}{
    J^{\frac{1}{2}-\epsilon}
}. \nonumber
\end{equation}
or, equivalently,
\begin{equation}
\left|
    (bg^J,q^M) - (bg,q^M)
\right| \le \frac{
    \| b \|_\infty W_{g,\epsilon}
}{
    M^{\frac{1}{2}-\epsilon}
}. \label{eq_tri1term_eps}
\end{equation}
since we have assumed that $J=J(M)\ge M$.

Taking together \eqref{eqTriang1}, \eqref{eq_tri2term_eps} and
\eqref{eq_tri1term_eps} we obtain
\begin{equation}
| (bg^J,q^M) - (bg,q) | \le \frac{
    \| b \|_\infty W_{g,\epsilon} + U_\epsilon
}{
    M^{\frac{1}{2}-\epsilon}
} \label{eqBasica}
\end{equation}
and it is immediate to combine the inequality \eqref{eqInicial} with
the bound in \eqref{eqBasica}. If we choose  $b=f$ in order to
control the first term on the right hand side of \eqref{eqInicial},
and $b=1$ in order to control the second term, we readily find that
\begin{equation}
| (f,\pi^{M,J}) - (f,\pi) | \le \frac{
    W_{f,g,\epsilon}
}{
    M^{\frac{1}{2}-\epsilon}
}, \label{eqProof_ineq1}
\end{equation}
where
$$
W_{f,g,\epsilon} = \frac{1}{(1,h)}\left[
    (1+\| f \|_\infty) W_{g,\epsilon} + 2U_\epsilon
\right] > 0
$$
is an a.s. finite random variable.

The proof for inequality \eqref{eq2Basic} is simpler. A triangle
inequality yields
\begin{equation}
| (f,\bar \pi^{M,J}) - (f,\pi) | \le | (f,\bar \pi^{M,J}) -
(f,\pi^{M,J}) | + | (f,\pi^{M,J}) - (f,\pi) | \label{eqTriang2}
\end{equation}
and \eqref{eqProof_ineq1} already provides an adequate bound for the
second term on the right hand side of  \eqref{eqTriang2}. For the
first term on the right hand side, we note that
\begin{equation}
(f,\bar \pi^{M,J}) = \frac{
    ( f[\varphi^M \circ g^J], q^M )
}{
    ( \varphi^M \circ g^J, q^M )
}, \label{eqNormalizada}
\end{equation}
where $\circ$ denotes composition, hence $(\varphi^M \circ
g^J)(\bftheta) = \varphi^M( g^J (\bftheta) )$. Taking together
\eqref{eqNormalizada} and the expression for $(f,\pi^{M,J})$ in
\eqref{eqNormalizada0} yields, by the same argument leading to
\eqref{eqInicial},
\begin{eqnarray}
| (f,\bar \pi^{M,J}) - (f,\pi^{M,J}) | &\le& \frac{
    | (f[\varphi^M\circ g^J], q^M) - (fg^J,q^M) |
}{
    (\varphi^M \circ g^J, q^M)
} \nonumber \\
&& + \frac{
    \| f \|_\infty | (\varphi^M\circ g^J, q^M) - (g^J,q^M) |
}{
    (\varphi^M \circ g^J, q^M)
}  \nonumber \\
&\le& a | (f[\varphi^M\circ g^J], q^M) - (fg^J,q^M) | \nonumber\\
&& + a \| f \|_\infty | (\varphi^M\circ g^J, q^M) - (g^J,q^M) |, \nonumber\\
\label{eqTriang3}
\end{eqnarray}
where the second inequality follows from the definition of
$\varphi^M$ in A1 and the bound $g^J\geq a^{-1}$ in A3.

In order to use \eqref{eqTriang3}, we look into errors of the form
$|(b[\varphi^M\circ g^J], q^M) - (bg^J,q^M)|$ for arbitrary $b \in
B(\sS)$. This turns out relatively straightforward since, from the
definition of $\varphi^M$ in A1,
\begin{eqnarray}
|(b[\varphi^M\circ g^J], q^M) - (bg^J,q^M)| &=& \nonumber\\
\left|
    \frac{1}{M}\sum_{r=1}^{M_T} b(\bftheta^{(i_r)}) \left[
        g^J(\bftheta^{(i_{M_T})}) - g^J(\bftheta^{(i_r)})
    \right]
\right| &\le& 2 a \|b\|_\infty\frac{M_T}{M},\nonumber\\
\label{eqFacil}
\end{eqnarray}
where the inequality follows from using uniform bound $g^J \leq a$
in A3. We can plug \eqref{eqFacil} into \eqref{eqTriang3} twice,
first choosing $b=f$ and then $b=1$, in order to control the two
terms in the triangle inequality. As a result, we arrive at the {\em
deterministic} bound
\begin{equation}
| (f,\bar \pi^{M,J}) - (f,\pi^{M,J}) | \le \frac{
    2a^2\|f\|_\infty M_T
}{
    M
} \le \frac{
    2a^2\|f\|_\infty
}{
    \sqrt{M}
}, \label{eqSegunda}
\end{equation}
where the second inequality follows from the assumption
$M_T\le\sqrt{M}$ in the statement of the Lemma.

Substituting \eqref{eqSegunda} and \eqref{eqProof_ineq1} back into
\eqref{eqTriang2} yields
\begin{equation}
| (f,\bar \pi^{M,J}) - (f,\pi^{M,J}) | \le \frac{
    W_{f,g,\epsilon} + 2a^2\| f \|_\infty
}{
    M^{\frac{1}{2}-\epsilon}
}, \label{eqProof_ineq2}
\end{equation}
which reduces to the inequality \eqref{eq2Basic} in the statement of
the Lemma, with $\bar W_{f,g,\epsilon} = W_{f,g,\epsilon} + 2a^2 \|
f \|_\infty > 0$ an a.s. finite random variable. $\qed$

\section{Proof of Lemma \ref{lmUniformL}} \label{apProofUniformL}

It can be proved \citep[Lemma 1]{Crisan2013nested} that for any $f \in
B(\mX)$
\begin{equation}
\label{eq_1} \sup_{\bftheta \in \sS} \| (f, \xi_{n,\bftheta}^J) -
(f, \xi_{n,\bftheta}) \|_p \leq \frac{c(f)}{\sqrt{J}},
\end{equation}
where $c(f)$ is a constant independent of $\bftheta$ and $J$. In
particular, there exists an a.s. finite non negative random variable
$U_{n,\bftheta,f,\epsilon}$, independent of $J$, such that
\begin{equation*}
\label{eq_7} | (f, \xi_{n,\bftheta}^J) - (f, \xi_{n,\bftheta}) | <
\frac{U_{n,\bftheta,f,\epsilon}}{J^{\frac{1}{2}-\epsilon}}
\end{equation*}
for any constant $0<\epsilon<\frac{1}{2}$ (see \citep[Lemma
4.1]{Crisan2011particle}).

Note that, while the constant $c(f)$ in (\ref{eq_1}) is independent
of $\bftheta$, the random variable $U_{n,\bftheta,f,\epsilon}$ is
not necessarily so. However, the inequality (\ref{eq_1}) holds for
every $\bftheta \in \sS$. Therefore $U_{n,\bftheta,f,\epsilon} \ge
0$ is a.s. finite for every $\bftheta \in \sS$, hence
\begin{equation*}
U_{n,f,\epsilon} := \sup_{\bftheta \in \sS}
U_{n,\bftheta,f,\epsilon} < \infty \quad \textrm{a.s.}
\end{equation*}
As a consequence, for any $f \in B(\mX)$,
\begin{equation}
\label{eq_2}\sup_{\bftheta \in \sS} | (f,\xi_{n,\bftheta}^J) -
(f,\xi_{n,\bftheta}) | \leq \sup_{\bftheta \in \sS}
\frac{U_{n,f,\bftheta,\epsilon}}{J^{\frac{1}{2}-\epsilon}} \leq
\frac{U_{n,f,\epsilon}}{J^{\frac{1}{2}-\epsilon}},
\end{equation}
where $U_{n,f,\epsilon}\ge 0$ is a.s. finite and independent of
$\bftheta$ and $J$.

Now, given the record of observations $\textbf{y}_{1:N}$ we need to
find error rates for the likelihood of $\bftheta$, namely for
$\lambda_N(\bftheta) = \prod_{n=1}^N (u_n^{\textbf{y}_n}, \xi_{n,\bftheta})$, where
$u_n^{\textbf{y}_n} \in B(\mX)$ and $\bftheta \in \sS$. Using the inequality (\ref{eq_2}) we obtain
\begin{equation}
(u_n^{\textbf{y}_n}, \xi_{n,\bftheta}) -
\frac{U_{n,u,\epsilon}}{J^{\frac{1}{2}-\epsilon}} \leq
(u_n^{\textbf{y}_n}, \xi_{n,\bftheta}^J) \leq (u_n^{\textbf{y}_n},
\xi_{n,\bftheta}) +
\frac{U_{n,u,\epsilon}}{J^{\frac{1}{2}-\epsilon}}
\label{eqDobleineq}
\end{equation}
a.s. for every $\bftheta \in \sS$ (where the random variables
$U_{n,u,\epsilon}$ is independent of $\bftheta$ and $J$, and a.s.
finite) and, since $(u_n^{\textbf{y}_n}, \xi_{n,\bftheta}^J) > 0$ by
assumption, Eq. \eqref{eqDobleineq} readily yields
\begin{eqnarray}
0 \vee \prod_{n=1}^N \left[ (u_n^{\textbf{y}_n}, \xi_{n,\bftheta}) -
\frac{U_{n,u,\epsilon}}{J^{\frac{1}{2}-\epsilon}} \right] &\leq&
\prod_{n=1}^N (u_n^{\textbf{y}_n}, \xi_{n,\bftheta}^J) \nonumber
\\ &\leq& \prod_{n=1}^N \left[ (u_n^{\textbf{y}_n},
\xi_{n,\bftheta}) +
\frac{U_{n,u,\epsilon}}{J^{\frac{1}{2}-\epsilon}}
\right], \nonumber\\
\label{eq_3}
\end{eqnarray}
where $a \vee b$ denotes the maximum between $a$ and $b$.

The term on the right hand side of \eqref{eq_3} can be decomposed as
\begin{eqnarray*}
\prod_{n=1}^N \left[ (u_n^{\textbf{y}_n}, \xi_{n,\bftheta}) +
\frac{U_{u,n,\epsilon}}{J^{\frac{1}{2}-\epsilon}} \right] = \left(
\prod_{n=1}^N (u_n^{\textbf{y}_n},
\xi_{n,\bftheta}) \right) + \\
\sum_{\alpha \in A^N} 
\prod_{n=1}^N (u_n^{\textbf{y}_n}, \xi_{n,\bftheta})^{\alpha_n}
\times \left( \frac{U_{u,n,\epsilon}}{J^{\frac{1}{2}-\epsilon}}
\right)^{1-\alpha_n},
\end{eqnarray*}
where $\alpha = (\alpha_1, \ldots, \alpha_n) \in \{0,1\}^N$ is a
multi-index of $0/1$ entries and $A^N = \{0,1\}^N \backslash
(1,\ldots,1)$ is the set of all such multi-indices excluding $(1,
..., 1)$. For every $\alpha \in A^N$, the factor $
V_{N,u,\alpha_n,\epsilon} = \prod_{n=1}^N (u_n^{\textbf{y}_n},
\xi_{n,\bftheta})^{\alpha_n} U_{n,u,\epsilon}^{1-\alpha_n} $ is a
random variable and, since $N$ is finite, $(u_{n}^{\textbf{y}_n},
\xi_{n,\bftheta}) \leq \| u_n^{\textbf{y}_n} \|_\infty < \infty$ and
$U_{n,u,\epsilon} < \infty$ a.s., it turns out that
\begin{equation*}
V_{N,u,\alpha_n,\epsilon} = \prod_{n=1}^N (u_n^{\textbf{y}_n},
\xi_{n,\bftheta})^{\alpha_n} U_{n,u,\epsilon}^{1-\alpha_n} < \infty
\quad \textrm{a.s.}
\end{equation*}
and, again, since $N < \infty$
\begin{equation*}
V_{N,u,\epsilon} = \sum_{\alpha_n \in A^N} V_{N,u,\alpha_n,\epsilon}
< \infty \quad \textrm{a.s.}
\end{equation*}
(a sum of a.s. finite random variables). Moreover, every $\alpha \in
A^N$ contains at least one $0$ entry, hence
\begin{equation}
\label{eq_4} \prod_{n=1}^N \left[ (u_n^{\textbf{y}_n}, \xi_{n,
\bftheta}) + \frac{U_{n,u,\epsilon}}{J^{\frac{1}{2}-\epsilon}}
\right] \leq \prod_{n=1}^N (u_n^{\textbf{y}_n}, \xi_{n,\bftheta}) +
\frac{V_{N,u,\epsilon}}{J^{\frac{1}{2}-\epsilon}}.
\end{equation}
By a similar argument, there exists an a.s. finite random variable
$\tilde{V}_{N,u,\epsilon}$ such that
\begin{equation}
\label{eq_5} \prod_{n=1}^N \left[ (u_n^{\textbf{y}_n}, \xi_{n,
\bftheta}) - \frac{U_{n,u,\epsilon}}{J^{\frac{1}{2}-\epsilon}}
\right] \geq \prod_{n=1}^N (u_n^{\textbf{y}_n}, \xi_{n,\bftheta}) -
\frac{\tilde{V}_{N,u,\epsilon}}{J^{\frac{1}{2}-\epsilon}}.
\end{equation}

Combining (\ref{eq_3}), (\ref{eq_4}) and (\ref{eq_5}), we obtain
\begin{eqnarray}
\label{eq_6} 0 \vee \left( \prod_{n=1}^N (u_n^{\textbf{y}_n},
\xi_{n, \bftheta}) -
\frac{\tilde{V}_{N,u,\epsilon}}{J^{\frac{1}{2}-\epsilon}} \right)
&\leq&
\prod_{n=1}^N (u_n^{\textbf{y}_n}, \xi_{n,\bftheta}^J) \\
\nonumber &\leq& \prod_{t=1}^T (u_n^{\textbf{y}_n},
\xi_{n,\bftheta}) +
\frac{V_{N,u,\epsilon}}{J^{\frac{1}{2}-\epsilon}}.
\end{eqnarray}
Finally, if we introduce
\begin{equation*}
W_{N,u,\epsilon} = V_{N,u,\epsilon} \vee \tilde{V}_{N,u,\epsilon} <
\infty \quad \textrm{a.s.},
\end{equation*}
then (\ref{eq_6}) yields
\begin{equation*}
\left| \prod_{n=1}^N (u_n^{\textbf{y}_n}, \xi_{n,\bftheta}^J) -
\prod_{n=1}^N (u_n^{\textbf{y}_n}, \xi_{n,\bftheta}) \right| \leq
\frac{W_{N,u,\epsilon}}{J^{\frac{1}{2}-\epsilon}},
\end{equation*}
where $0 \leq W_{N,u,\epsilon} < \infty$ a.s.

\bibliographystyle{spbasic}         
\bibliography{biblio}

\begin{thebibliography}{26}
\providecommand{\natexlab}[1]{#1}
\providecommand{\url}[1]{{#1}}
\providecommand{\urlprefix}{URL }
\expandafter\ifx\csname urlstyle\endcsname\relax
  \providecommand{\doi}[1]{DOI~\discretionary{}{}{}#1}\else
  \providecommand{\doi}{DOI~\discretionary{}{}{}\begingroup
  \urlstyle{rm}\Url}\fi
\providecommand{\eprint}[2][]{\url{#2}}

\bibitem[{Andrieu et~al(2010)Andrieu, Doucet, and Holenstein}]{Andrieu2010}
Andrieu C, Doucet A, Holenstein R (2010) Particle {M}arkov chain {M}onte
  {C}arlo methods. Journal of the Royal Statistical Society: Series B
  (Statistical Methodology) 72(3):269--342

\bibitem[{Bain and Crisan(2008)}]{Bain2008}
Bain A, Crisan D (2008) Fundamentals of stochastic filtering, vol~60. Springer
  Verlag

\bibitem[{Bengtsson et~al(2008)Bengtsson, Bickel, and Li}]{Bengtsson08}
Bengtsson T, Bickel P, Li B (2008) Curse of dimensionality revisited: Collapse
  of particle filter in very large scale systems. Probability and statistics:
  Essay in honour of David A Freedman 2:316--334

\bibitem[{Boys et~al(2008)Boys, Wilkinson, and Kirkwood}]{Boys08}
Boys RJ, Wilkinson DJ, Kirkwood TBL (2008) Bayesian inference for a discretely
  observed stochastic kinetic model. Statistics and Computing 18(2):125--135

\bibitem[{Capp\'e et~al(2004)Capp\'e, Guillin, Marin, and Robert}]{Cappe04}
Capp\'e O, Guillin A, Marin JM, Robert CP (2004) Population {M}onte {C}arlo.
  Computational and Graphical Statistics 13(4):907--929

\bibitem[{Capp\'e et~al(2008)Capp\'e, Douc, Guillin, Marin, and
  Robert}]{Cappe08}
Capp\'e O, Douc R, Guillin A, Marin JM, Robert CP (2008) Adaptive importance
  sampling in general mixture classes. Statistics and Computing 18(4):447--459

\bibitem[{Crisan and M\'{\i}guez(2011)}]{Crisan2011particle}
Crisan D, M\'{\i}guez J (2011) Particle approximation of the filtering density
  for state-space {M}arkov models in discrete time. arXiv preprint
  arXiv:11115866

\bibitem[{Crisan and M\'{\i}guez(2013)}]{Crisan2013nested}
Crisan D, M\'{\i}guez J (2013) Nested particle filters for online parameter
  estimation in discrete-time state-space {M}arkov models. arXiv preprint
  arXiv:13081883

\bibitem[{Doucet et~al(2000)Doucet, Godsill, and Andrieu}]{Doucet00}
Doucet A, Godsill S, Andrieu C (2000) On sequential {M}onte {C}arlo {S}ampling
  methods for {B}ayesian filtering. Statistics and Computing 10(3):197--208

\bibitem[{Doucet et~al(2001)Doucet, De~Freitas, and Gordon}]{Doucet2001}
Doucet A, De~Freitas N, Gordon N (2001) Sequential {M}onte {C}arlo methods in
  practice. Springer Verlag

\bibitem[{Gillespie(1977)}]{Gillespie77}
Gillespie DT (1977) Exact stochastic simulation of coupled chemical reactions.
  The Journal of Physical Chemistry 81(25):2340--2361

\bibitem[{Golightly and Wilkinson(2005)}]{Golightly2005}
Golightly A, Wilkinson DJ (2005) Bayesian inference for stochastic kinetic
  models using a diffusion approximation. Biometrics 61(3):781--788

\bibitem[{Golightly and Wilkinson(2011)}]{Golightly2011}
Golightly A, Wilkinson DJ (2011) Bayesian parameter inference for stochastic
  biochemical network models using particle {M}arkov chain {M}onte {C}arlo.
  Interface Focus 1(6):807--820

\bibitem[{Gordon et~al(1993)Gordon, Salmond, and Smith}]{Gordon1993}
Gordon NJ, Salmond DJ, Smith AF (1993) Novel approach to
  nonlinear/non-{G}aussian {B}ayesian state estimation. In: IEE Proceedings F
  (Radar and Signal Processing), IET, vol 140, pp 107--113

\bibitem[{Kilbinger(2010)}]{Kilbinger2010}
Kilbinger Mea (2010) Bayesian model comparison in cosmology with population
  {M}onte {C}arlo. Royal astronomical society

\bibitem[{Kilbinger(2012)}]{Kilbinger2012}
Kilbinger Mea (2012) Cosmo{PMC}: Cosmology population {M}onte {C}arlo. arXiv
  preprint arXiv:11010950v3

\bibitem[{Koblents and M\'{\i}guez(2013{\natexlab{a}})}]{Koblents2013b}
Koblents E, M\'{\i}guez J (2013{\natexlab{a}}) A population {M}onte {C}arlo
  scheme for computational inference in high dimensional spaces. ICASSP

\bibitem[{Koblents and M\'{\i}guez(2013{\natexlab{b}})}]{Koblents2013a}
Koblents E, M\'{\i}guez J (2013{\natexlab{b}}) A population {M}onte {C}arlo
  scheme with transformed weights and its application to stochastic kinetic
  models. Statistics and Computing pp 1--19, \doi{10.1007/s11222-013-9440-2},
  \urlprefix\url{http://dx.doi.org/10.1007/s11222-013-9440-2}

\bibitem[{Koblents and M\'{\i}guez(2013{\natexlab{c}})}]{Koblents2013c}
Koblents E, M\'{\i}guez J (2013{\natexlab{c}}) Robust mixture population
  {M}onte {C}arlo scheme with adaptation of the number of components. EUSIPCO

\bibitem[{Lewis and Bridle(2002)}]{Lewis2002}
Lewis A, Bridle S (2002) Cosmological parameters from cmb and other data: a
  {M}onte {C}arlo approach. Phys Rev D66:103,511, \eprint{astro-ph/0205436}

\bibitem[{Milner et~al(2013)Milner, Gillespie, and Wilkinson}]{Milner2013}
Milner P, Gillespie C, Wilkinson D (2013) Moment closure based parameter
  inference of stochastic kinetic models. Statistics and Computing pp 1--9

\bibitem[{Robert and Casella(2004)}]{Robert04}
Robert CP, Casella G (2004) {M}onte {C}arlo Statistical Methods. Springer

\bibitem[{Volterra(1926)}]{Volterra26}
Volterra V (1926) Fluctuations in the abundance of a species considered
  mathematically. Nature 118:558--560

\bibitem[{Wilkinson(2011{\natexlab{a}})}]{Wilkinson2011paper}
Wilkinson D (2011{\natexlab{a}}) Parameter inference for stochastic kinetic
  models of bacterial gene regulation: A {B}ayesian approach to systems
  biology. (with discussion), in Bayesian Statistics 9

\bibitem[{Wilkinson(2011{\natexlab{b}})}]{Wilkinson2011}
Wilkinson D (2011{\natexlab{b}}) Stochastic modelling for systems biology,
  vol~44. CRC press

\bibitem[{Wraith(2009)}]{Wraith2009}
Wraith Dea (2009) Estimation of cosmological parameters using adaptive
  importance sampling. arXiv preprint arXiv:09030837v1

\end{thebibliography}


\end{document}